\documentstyle[aps,preprint,eqsecnum]{revtex}
\begin{document}
\draft
\title{Bosonization Based on Bethe Ansatz Equations and\\
Spin-Charge Separation in the Hubbard Model with Finite $U$}
\author{Yue Yu $^1$, Huan-Xiong Yang $^{2,1}$ and Yong-Shi Wu $^3$}
\address{1. Institute of Theoretical Physics, Chinese \\
Academy of Sciences, Beijing 100080, P. R. China}
\address{2. Department of Physics, Zhejiang University, \\
Hangzhou, 310027, P. R. China}
\address{3. Department of Physics, University of Utah, \\
Salt Lake City, UT 84112, U.S.A.}
\date{\today}
\maketitle

\begin{abstract}
We develop a bosonization approach for one-dimensional models based on Bethe
ansatz equations. The operator formalism of the exact soluble models in the
low energy limit provides a systematic method to calculate the asymptotic
correlation functions. As examples with and without internal degrees of
freedom, the Calogero-Sutherland (C-S) model and the repulsive Hubbard model
are considered respectively. We verify that the low energy behavior of the
C-S model is controlled by two classes of $c=1$ conformal field theories,
depending on whether the C-S interactions are among bosons or among
fermions. For the Hubbard model, we show the explicit charge-spin separation
at low energy for arbitrary $U>0$. The low energy behavior of the system is
described by the (semi-) direct product of two independent Virasoro algebras
with $c=1$.
\end{abstract}

\pacs{}

%%%%%%%%%%%%%%%%%%%%%%%

%%%%%%%%%%%%%%%%%%%%%%%

%\receipt{}

\section{Introduction}

Recently it has attracted much attention to understand strongly correlated
fermion systems in low dimensions ($d\leq 2$). Thanks to the special
behaviors of one-dimensional (1-d) systems, we have better knowledge for
them than those in higher dimensions, with the help from some
non-perturbative methods. There is a number of 1-d systems called exactly
soluble, whose spectra are exactly given by the Bethe ansatz equations.
However, it is often difficult to calculate the correlation functions by
using the Bethe ansatz. On the other hand, the bosonization approach
provides a useful tool to compute the asymptotic correlation functions \cite
{TomMatt} in certain cases. Therefore, it is important to understand the
inter-relationship between exact solutions and bosonization. According to
Haldane \cite{Hald1}, the Bethe ansatz soluble models can be characterized
by 1-d Luttinger liquids. Moreover, after renormalization, only forward
scatterings are important in the Luttinger liquids. These imply that there
is a harmonic fluid description \cite{Hald2} to exact soluble models in the
low energy limit, i.e., it should be possible to bosonize these models, at
least in the low energy regime. However, how to do bosonization, especially
for the case of the systems with internal degrees of freedom, has never been
shown explicitly before in the literature. In this paper we will show how to
bosonize the exact soluble models at low energy by starting from the Bethe
ansatz.. The bonus of doing so is a short cut for deriving the low
temperature thermodynamics and static correlation functions. We will take
two nontrivial models, the Calogero-Sutherland (C-S) model and the Hubbard
model as examples.

The C-S model \cite{Ca,Su} not only presents a beautiful example for soluble
1-d many-body model with long range interactions, but is also closely
related to physical phenomena such as the edge excitations in the fractional
quantum Hall effect \cite{Hald3}\cite{Yue}. It also provides a realization
of some basic physical concept such as exclusion statistics \cite{Hald4,WuBe}%
. The correlation functions of the C-S model can be exactly calculated in a
wide range of the coupling constant \cite{Ha,SHF}. Also, Ha \cite{Ha}
proposed a bosonized anyon effective theory at low energy, and found that
the asymptotic correlation functions of the effective theory agree with
those of exact results. This raises a question of whether there is an
explicit way to derive the bosonization and show this agreement. In our
recent work \cite{Wu}, a bosonization approach was developed for 1-d gas
with exclusion statistics, which can be modified to deal with this problem.
We will apply this approach to the bosonization of the C-S model, as a
warm-up for treating models with internal degrees of freedom such as the
Hubbard model. A by-product of this exercise is the proof that at low energy
the bosonic and fermionic C-S models give rise to two different $c=1$
conformal field theories (CFT's), because of different selection rules for
the quantum numbers. To our knowledge, this result is new in the literature.
We also present the correlation functions respectively in each case.

The Hubbard model is known to have many remarkable properties, and has
attracted a lot of attentions since the two dimensional Hubbard model was
believed to be the right model for the high $T_c$-superconductivity about
ten years ago. Although the 1-d Hubbard model we studied here may not
directly relate to two-dimensional physics, there are still good reasons to
investigate it for gaining insights into high $T_c$-superconductivity. After
Lieb-Wu's Bethe ansatz solution of the 1-d Hubbard model \cite{Lieb}, a lot
of efforts have been spent in this model. We list here some relevant
developments. As is well-known, there may be string solutions with complex
rapidities in some Bethe ansatz soluble models, which may affect the low
energy physics of the system \cite{HKKW}. Fortunately, for the Hubbard
model, it has been shown that the complex rapidities offer only the states
in the upper Hubbard band and, therefore, do not contribute to the low
energy physics \cite{Woy}. So we are allowed to consider only the real
rapidities in our treatment. Spin-charge separation is a fascinating
property of the Hubbard model. It has been shown that in the strong-coupling
limit, the ground-state wave function given by Lieb-Wu's solution is indeed
of the spin-charge separation form \cite{OS}. It was also checked from the
exclusion statistics point of view that the low-lying states are also of the
spin-charge separation form for large $U$ \cite{HKKW}. Ren and Anderson \cite
{RA} assumed the separation for arbitrary coupling in the low-energy limit
and proposed a bosonized effective theory of the model. The critical
exponents of the effective theory agreed with the numerical calculations 
\cite{OS,So} and the phase shift argument \cite{RA}. The correlation
functions of the Hubbard model were also discussed by finite-size scaling
considerations \cite{W-FK}. It has been proven that there are two
independent $c=1$ Virasoro algebras which describe the critical theory of
the Hubbard model. It was also asserted that the low energy behaviors of the
system at less than half filling could be characterized by a (semi-)direct
product of two Virasoro algebras \cite{W-FK}.

In this paper, we consider the repulsive Hubbard model at less than half
filling. The low energy effective theory of the Hubbard model is not Lorentz
invariant in $1+1$ dimensions, since there are two different Fermi vectors.
This generally leads to the gap-less charge- and spin-density waves having
different velocities. In this sense, the spin-charge separation is expected
to appear not only at large $U$ but also at low energy at arbitrary
coupling, as has been assumed in \cite{RA}. Starting form the Bethe ansatz
solutions, we will explicitly demonstrate this property in the low
temperature thermodynamic limit. We show that the free energy of the model
at the low temperature is of the form $F(T)/L=F(0)/L-\frac{\pi T^2}{6v_c}-%
\frac{\pi T^2}{6v_s}$, which implies that the critical behavior of the
system can be characterized by the direct product of two independent $c=1$
Virasoro algebras and coincides with the results of finite-size scaling at $%
T=0$ \cite{W-FK}. Meanwhile, by generalizing the bosonization approach
(without internal degrees of freedom) to the Hubbard model which
incorporates spin degrees of freedom, we arrive at an effective theory by
bosonizing the thermodynamic limit of the original theory, resulting in the
bosonized effective theory proposed by Ren and Anderson. Then the
calculations of the asymptotic correlation functions become systematic.

This paper is organized as follows: In the next section, we present
bosonization of the C-S model. In Sec. III, some relevant results about the
Hubbard model are reviewed. Then in Sec. IV, the thermodynamic potential of
the Hubbard model (and then free energy) at low temperature is derived from
the Bethe Ansatz equations and the spin-charge separation at low energy is
demonstrated. In Sec. V the bosonization of the Hubbard model is given, and
the single-particle correlation functions are systematically calculated for
various excitations. The last section is devoted to the conclusions.

\section{Calogero-Sutherland Model}

The C-S model we consider here is described by the $N$-body Hamiltonian 
\begin{equation}
H=-\sum_{i=1}^N \frac{\partial^2}{\partial x_i^2} +\sum_{i<j}U(x_i-x_j),
\label{hamil}
\end{equation}
with 
\begin{equation}
U(x)=g\sum_{n=-\infty}^{\infty}(x+nL)^{-2} =g\frac{\pi^2}{L^2}\sin^{-2}(%
\frac{ \pi x}{L}),  \label{poten}
\end{equation}
where $L$ is the size of 1-d ring. As $L\to \infty$, the interaction
potential (\ref{poten}) $U(x)\to g/x^2$. The ground state is given by \cite
{Su} 
\begin{equation}
\psi_{B,\lambda}=\prod_{i<j} |\sin\pi(x_i-x_j)/L|^\lambda,
\end{equation}
or 
\begin{equation}
\psi_{F,\lambda}=\prod_{i<j}(z_i-z_j)^\lambda \prod_k z_k^{-\lambda(n-1)/2},
\end{equation}
depending on the statistics of the particles, bosons or fermions. Here $%
z=\exp(i2\pi x/L)$ and $\lambda=[(1+2g)^{1/2}+1]/2$. It is obvious that for $%
\lambda=1(g=0)$ the former is the ground state of hard-core bosons and the
latter that of free fermions.

Moreover, the model can be exactly solved by using the asymptotic Bethe
ansatz \cite{Su} or the Jack Polynomials \cite{Ha}. The periodic boundary
conditions give rise to the spectrum determined by the following equations
for pseudomomenta $k_i$: 
\begin{equation}
Lk_i=2\pi I_i+\pi(1-\lambda) \sum_{j<i}{\rm sgn}(k_j-k_i),  \label{ABA}
\end{equation}
where \cite{comm1} 
\begin{equation}
I_i=\Biggl\{ {\ 
\begin{array}{ll}
(N+1)/2~ {\rm mod}(1), & {\rm for~fermion~case}, \\ 
\;\; {\rm integer}, & {\rm for~boson~case}
\end{array}
}  \label{select}
\end{equation}
These selection rules have been discussed by Kawakami and Yang, and for
bosons and fermions the behavior of the momentum distribution is quite
different \cite{KaYa}. We will see that it is the difference in the
selection rules that gives rise to two different classes of $c=1$ CFT for
low energy effective field theory. The total energy $E$ and the total
momentum $P$ are, respectively, 
\begin{equation}
E=\sum_i k_i^2,~~~P=\sum_i k_i=\sum_i p_i(k).  \label{TEP}
\end{equation}
where $p_i(k)=2\pi I_i/L$

The thermodynamics of the theory has been fully discussed by Sutherland \cite
{Su}. Here we would like to examine its relations to CFT, Luttinger liquid
and bosonization. Some relevant topics have been partially considered by
several authors before \cite{KaYa,Ha,Car,Iso}. The thermodynamics of the C-S
model can also be formulated in terms of an ideal excluson gas (IEG) \cite
{Hald4,WuBe}. The thermodynamic potential is known to be given by \cite
{Su,WuBe} 
\begin{equation}
\frac{\Omega}{L}=-\frac{T}{2\pi} \int_{-\infty}^\infty dk \ln(1+w(k,T)^{-1}),
\label{Omega}
\end{equation}
with the function $w(k,T)\equiv \rho_{a}(k)/\rho(k)$, where $\rho_{a}(k)$
and $\rho(k)$ are the hole and particle densities, satisfying 
\begin{equation}
w(k,T)^{\lambda} [1+w(k,T)]^{1-\lambda} =e^{(k^2-\mu)/T}.  \label{forW}
\end{equation}
The particle density is determined by 
\begin{equation}
\rho(k,T)(1+w(k,T))=\frac{1}{2\pi} -(\lambda-1)\rho(k,T).  \label{rhoCS}
\end{equation}

In the ground state, there is a (pseudo-)Fermi surface $k_{F}$, such that $%
\rho(k)=1/2\pi\lambda$ for $|k|<k_{F}$ and $\rho(k)=0$ for $|k|>k_{F}$. Then
the Fermi momentum is given by $k_F =\pi \lambda \bar{d}_{0}$, and the
ground state energy and momentum by $E_0/L=\pi^2\lambda^2 \bar{d}_{0}^{3}/3$%
, $P_{0}=0$, where $\bar{d}_{0}=N_{0}/L$.

Now let us examine possible excitations in the model. First there are
density fluctuations due to particle-hole excitations, i.e., the sound waves
with velocity $v_{s} =v_{F}\equiv 2k_{F}$ (see below). Moreover, by adding
extra $M$ particles to the ground state one can create particle excitations,
and by Galileo boost a persistent current. It is easy to verify that the
velocities of these three classes of elementary excitations in the model
satisfy a fundamental relation, $v_{s}=\sqrt{v_{N}v_{J}}$, that Haldane
years ago used to characterize the Luttinger liquid \cite{Hald5}. Indeed,
shifting $N_0$ to $N=N_0+M$, the change in the ground state energy is $%
\delta_1 E_0=\pi(\lambda k_F)M^2$, while a persistent current, created by
the boost of the Fermi sea $k\to k+\pi J/N_0$, leads to the energy shift $%
\delta_2E_0 =\pi(k_F/\lambda)J^2$. Therefore the total changes in energy and
in momentum, due to charge and current excitations, are 
\begin{eqnarray}
\delta E_0&=&(\pi/2L)(v_N M^2+v_J J^2),  \nonumber \\
\delta P_0&=&\pi(\bar{d}_0+M/L)J,  \label{CSDE}
\end{eqnarray}
with 
\begin{equation}
v_N=v_s\lambda, \;\; v_J=v_{s}/\lambda, \;\; v_{s} = \sqrt{v_{N}v_{J}}.
\label{Velo}
\end{equation}
These coincide with the well-known relations \cite{Hald5} in the Luttinger
liquid theory, if we identify $\lambda$ with Haldane's controlling parameter 
$\exp(-2\varphi)$. It is easy to check that the selection rule (\ref{select}%
) can be rewritten as \cite{KaYa} 
\begin{equation}
J=\Biggl\{ {\ 
\begin{array}{ll}
M {\rm mod}(2), & {\rm for~fermion~case}, \\ 
\;\; {\rm even}, & {\rm for~boson~case}
\end{array}
}  \label{selectE}
\end{equation}

It has been claimed from finite-size scaling \cite{KaYa} that the low energy
behavior of the C-S model is controlled by $c=1$ CFT. A review of this point
will help to understand the behavior of the Hubbard model. The central
charge $c$ of a conformal invariant system is related to thermodynamic
observable by finite-size scaling \cite{CA}. Generally, the free energy of
the system is given by 
\begin{equation}
\frac{F(T)}{L}-\frac{F(0)}{L} =-\frac{\pi T^2 c}{6v_s}.  \label{FRE}
\end{equation}
For the C-S model, we consider the low energy expansion of the thermodynamic
potential (\ref{Omega}). Following Yang and Yang \cite{YangYang,Su}, we
introduce the dressed energy $\epsilon (k,T)$ by writing 
\begin{equation}
w(k,T)=e^{\epsilon (k,T)/T}.  \label{dressE}
\end{equation}
And (\ref{forW}) reads 
\begin{equation}
\epsilon(k,T)=k^2-\mu-T(1-\lambda) \ln(1+e^{-\epsilon(k,T)/T}).  \label{et}
\end{equation}
Because there is no singularity in $\epsilon(k,T)$ at $T=0$, the zero
temperature dressed energy is given by 
\begin{equation}
\epsilon(k)=\Biggl\{ {\ 
\begin{array}{ll}
(k^2-k^2_F)/\lambda , & |k|<k_F, \\ 
\;\; k^2-k_F^2, & |k|>k_F.
\end{array}
}  \label{EXP2}
\end{equation}

At low energy, one can consider only the excitation around the Fermi surface
and the thermodynamic potential is given by \cite{WuYue} 
\begin{equation}
\displaystyle \frac{\Omega(T)}{L} =\frac{1}{2\pi}\int_{-k_F}^{k_F} dk
\epsilon(k)-\frac{T}{\pi}\int_{-\delta}^{\delta} d\delta p_F
\ln(1+e^{-|\epsilon(k(p),T)|/T}),  \label{CSFRE1}
\end{equation}
where $\delta$ is a cut-off and the first term on the right hand side of the
last equality is recognized as $\Omega(0)/L$. $\delta p_F$ is the
deformation of the Fermi surface, i.e., $\delta p_F/2\pi=\rho(k_F)\delta k_F$%
. It is known that $\epsilon(k)=v_s\delta p_F$ near the Fermi point $p_F$.
After a little calculation, (\ref{CSFRE1}) reads 
\begin{equation}
\frac{\Omega(T)}{L}-\frac{\Omega(0)}{L} =-\frac{\pi T^2}{6v_s},  \label{CE1}
\end{equation}
which implies that the theory is actually cut-off independent at low
temperature. Notice that $F=\Omega-\mu N$. Only the particle-hole
excitations near the Fermi surface contribute to thermal excitations,
leading to $N(T)-N(0)=0$; this can be checked by an explicit calculation in
terms of the definition (\ref{rhoCS}) of $\rho(k,T)$. Thus, we have 
\begin{equation}
\frac{F(T)}{L}-\frac{F(0)}{L} =\frac{\Omega(T)}{L}-\frac{\Omega(0)}{L} =-%
\frac{\pi T^2}{6v_s}.  \label{FREE}
\end{equation}
Comparing this to (\ref{FRE}), we see $c=1$ for the C-S model.

The central charge of a conformal invariant system can also be given by the
finite-size scaling in the spatial direction \cite{CA}. In Appendix A, we
confirm $c=1$ for the C-S model from such a calculation, with some
subtleties not noticed before in the literature. Now let us work out
explicitly the effective theory at low energy with $c=1$ conformal
invariance. For this purpose, we will develop a bosonization approach. Our
starting point is the observation that the grand partition function $Z_G$,
corresponding to the thermodynamic potential (\ref{Omega}), is of the form
of that for a system of ideal fermions with a complicated, $T$-dependent
energy dispersion given by the dressed energy: 
\begin{equation}
Z_G=\prod_k(1+e^{-\epsilon (k,T)/T}).  \label{PFCS}
\end{equation}
This fermion representation is not very useful, because of the implicit $T$%
-dependence in the dressed energy. Nevertheless, as we have seen in ref. 
\cite{Wu,WuYue}, in the low-$T$ limit the dressed energy is effectively $T$%
-independent: Namely we have $\epsilon (k,T) =\epsilon (k)+O(Te^{-|\epsilon
(k)|/T})$, as a result of (\ref{et}). Hence, the grand partition function
can be obtained from an effective Hamiltonian, given by 
\begin{equation}
H_{{\rm eff}}=\sum_k {\epsilon}(k) \;c_k^\dagger c_k,  \label{EHCS}
\end{equation}
where $c_k^\dagger$ are fermion creation operators.

Another simplification in the low-$T$ limit is that we need to consider only
low-energy excitations near the right and left Fermi points $|k|= \pm k_F$,
which are completely decoupled and have a linearized energy dispersion 
\begin{equation}
\epsilon_\pm (k)=\Biggl\{ {\ 
\begin{array}{ll}
\pm v_F(k\mp k_F), \;\;\; & |k|>k_F, \\ 
\pm v_F (k\mp k_F)/\lambda, & |k|<k _F.
\end{array}
}
\end{equation}
We note the `refraction' at $k=\pm k_F$. In spite of this peculiarity, we
have succeeded in bosonizing the effective Hamiltonian as follows \cite{Wu}:
The density fluctuation operator at $k\sim k_F$ is constructed as follows 
\begin{eqnarray}
&&\rho_q^{(+)}=\displaystyle \sum_{k>k_F} :c^\dagger_{k+q}c_k: + %
\displaystyle \sum_{k<k_F-\lambda q} :c^\dagger_{k+\lambda q}c_k:  \nonumber
\\
&&+ \displaystyle \sum_{k_F-\lambda q< k < k_F} :c^\dagger_{\frac{k-k_F}{%
\lambda}+k_F+q}c_k:
\end{eqnarray}
for $q>0$. %and $D_+=( k_F, k_F-\lambda q)$. 
A similar density operator $\rho_q^{(-)}$ can also be defined at $k\sim -k_F$%
, 
\begin{eqnarray}
&&\rho_q^{(-)}=\displaystyle \sum_{k<-k_F} :c^\dagger_{k-q}c_k: + %
\displaystyle \sum_{k>-k_F+\lambda q} :c^\dagger_{k-\lambda q}c_k:  \nonumber
\\
&&+ \displaystyle \sum_{-k_F+\lambda q> k >- k_F} :c^\dagger_{\frac{k+k_F}{%
\lambda}-k_F-q}c_k:
\end{eqnarray}
Within the Tomonaga approximation \cite{comm2}, in which commutators are
taken to be their ground-state expectation value, we obtain 
\begin{equation}
[\rho_q^{(\pm)},\rho_{q^{\prime}}^{(\pm)\dagger}]= \frac{L}{2\pi} q\,
\delta_{q,q^{\prime}},\;\ [H_{\pm},\rho_{q}^{(\pm)}]= \pm v_Fq\rho_q^{(\pm)},
\label{density}
\end{equation}
which describe 1-d free phonons with the sound velocity $v_{s}=v_F$.
Introducing normalized boson annihilation operators $b_q=\sqrt{2\pi/ qL}%
\,\rho_q^{(+)}$ and $\tilde{b}_q=\sqrt{2\pi/qL}\,{\rho}_q^{(-)\dagger}$, the
bosonized Hamiltonian satisfying (\ref{density}) is given by 
\begin{equation}
H_B=v_s\{ \sum_{q>0}q(b_q^\dagger b_q +\tilde{b}_q^\dagger \tilde{b}_q) +%
\frac{1}{2}\frac{\pi}{L}[\lambda M^2 +\frac{1}{\lambda} J^2] \}.
\label{bosonH}
\end{equation}

The bosonized total momentum operator, corresponding to the fermionized $%
P=\sum_{k} p(k)\, c_{k}^{\dagger} c_{k}$, is 
\begin{equation}  \label{bosonP}
P=\displaystyle\sum_{q>0}q(b_q^\dagger b_q- \tilde{b}_q^\dagger \tilde{b}_q)
+ \pi (\bar{d}_0+M/L)\,J. \\
\end{equation}
In the coordinate-space formulation, the normalized density field $\rho(x)$
is given by $\rho(x)=\rho_R(x)+\rho_L(x)$: 
\begin{equation}
\rho_R(x)=\frac{M_R}{L}+ \displaystyle\sum_{q>0}\sqrt{q/2\pi L\lambda}
(e^{iqx}b_q+e^{-iqx}b_q^\dagger),  \label{rhofield}
\end{equation}
and $\rho_L(x)$ is similarly constructed from $\tilde{b}_q$ and $\tilde{b}%
_q^\dagger$, 
\begin{equation}
\rho_L(x)=\frac{M_L}{L}+ \displaystyle\sum_{q>0}\sqrt{q/2\pi L\lambda}
(e^{-iqx}\tilde{b}_q+e^{iqx}\tilde{b}_q^\dagger),
\end{equation}
where $M_{R,L}$ are given by $M=M_R+M_L$. The boson field $\phi(x)$, which
is conjugated to $\rho(x)$ and satisfies $[\phi(x),\rho(x^{\prime})]=i%
\delta(x-x^{\prime})$, is $\phi(x)=\phi_R(x)+\phi_L(x)$ with 
\begin{eqnarray}
\phi_R(x)&=& \frac{\phi_0}{2}+\frac{\pi J_Rx}{L}+i \displaystyle \sum_{q>0}%
\sqrt{\pi\lambda/ 2qL}(e^{iqx}b_q-e^{-iqx}b_q^\dagger),  \nonumber \\
\phi_L(x)&=& \frac{\phi_0}{2}+\frac{\pi J_Lx}{L}+i \displaystyle \sum_{q>0}%
\sqrt{\pi\lambda/ 2qL}(e^{-iqx}\tilde{b}_q-e^{iqx}\tilde{b}_q^\dagger), 
\nonumber \\
&&
\end{eqnarray}
with $J=J_R+J_L$. Here $M$ and $J$ are operators with integer eigenvalues
obeying the selection rule ({\ref{selectE}), and $\phi_0$ is an angular
variable conjugated to $M$: $[\phi_0,M]=i$. The Hamiltonian (\ref{bosonH}%
)becomes 
\begin{equation}
H_B =\frac{v_s}{2\pi} \int_0^Ldx\; [\Pi(x)^2+(\partial_xX(x))^2],
\label{fieldH}
\end{equation}
where $\Pi(x)=\pi\lambda^{1/2}\rho(x)$ and $X(x)=\lambda^{-1/2}\phi(x)$.
With $X(x,t)=e^{iHt}X(x)e^{-iHt}$, the Lagrangian density reads 
\begin{equation}
{\cal L}=\frac{v_s}{2\pi}\,\partial_\alpha X(x,t)\,\partial^\alpha X(x,t).
\end{equation}
}

We recognize in no ambiguity that ${\cal {L}}$ is the Lagrangian of a $c=1$
CFT \cite{CFT}. Since $\phi_{0}$ is an angular variable, there is a hidden
invariance in the theory under $\phi\to\phi+2\pi$. The field $X$ is thus
said to be ``compactified'' on a circle, with a radius that is determined by
the coupling constant: 
\begin{equation}
X\sim X+2\pi R,\;\;\; R^2=1/\lambda.
\end{equation}
(Incidentally, $\lambda$ can be identified with the exclusion statistics
parameter \cite{Wu} of the quasi-particle excitations in the C-S model.)
States $V[X]|0\rangle$ or operators $V[X]$ are {\it allowed} only if they
respect this invariance, so the quantum numbers of quasiparticles are
strongly constrained. For the boson case in (\ref{selectE}), the zero-mode
term in (\ref{bosonH}) can be written as $(\pi/2L)[ M^2/R^2+4R^2D^2]$, where 
$D=J/2=M {\rm mod} (1)$ are integers. This means that the Hamiltonian (\ref
{bosonH}) in this case has duality 
\begin{equation}
R \leftrightarrow 1/2R,\hspace{.2in} M \leftrightarrow D,
\end{equation}
which is just the famous duality in the usual $c=1$ CFT \cite{CFT}. For the
fermion case in (\ref{selectE}), the duality of (\ref{bosonH}) reads 
\begin{equation}
R\leftrightarrow 1/R,\hspace{.2in} M\leftrightarrow J,
\end{equation}
which corresponds to the so called excluson particle-hole duality mentioned
in \cite{WuBe,NaWi} 
\begin{equation}
\lambda\leftrightarrow 1/\lambda.
\end{equation}
Moreover, the partition function in the fermion case (in the low-$T$ limit)
can be rewritten as $Z=Tr_{{\cal H}}[q^{L^R_0} \bar{q}^{L^L_0}]$, where $%
q=e^{iv_s/T}$. The zeroth generators of the Virasoro algebra are 
\begin{eqnarray}
L^{R,L}_0 &=&v_s^{-1}H_{R,L}+\frac{\pi}{4L} [J R\mp M /R]^2,  \nonumber \\
H_R~&=& v_s\sum_{q>0} qb^\dagger _qb_q ,\hspace{.2in} H_L~= v_s\sum_{q>0} q%
\tilde{b}^\dagger _q\tilde{b}_q.
\end{eqnarray}
The constraint $J=M\,{\rm mod}\;(2)$ makes the spectrum and duality relation
different from those in the $c=1$ CFT compactified on a circle. Indeed, the
fermion case corresponds to the $c=1$ CFT compactified on an interval (or an
orbifold) $S^1/Z_2$; see \cite{CFT} for more details.

To show the difference between two CFT's, we check the primary fields in
each theory. The primary fields of both CFT's are given by 
\begin{equation}
\phi_{M,J}(x)\sim :e^{i(M\lambda^{1/2} +J/\lambda^{1/2})X_R(x)}
e^{i(M\lambda^{1/2} -J/\lambda^{1/2})X_L(x)}:.
\end{equation}
The charge-1 operators correspond to $M=1$. Therefore, charge-1 operators
are given by $M=1$ and $J=2m$ with integer $m$ for the boson case, while $%
J=2m+1$ for the fermion case. Hence, the charge-1 primary fields in the
boson case are bosons, and those in the fermion case are fermions: 
\begin{eqnarray}
&&\Psi^\dagger_B(x,t)=\rho(x)^{1/2} \sum_{m=-\infty}^{\infty} e^{iO_m}
:e^{i(\lambda^{1/2} + 2m/\lambda^{1/2})X_{R}(x_-)}:  \nonumber \\
&& :e^{i(\lambda^{1/2}-2m/\lambda^{1/2})X_{L}(x_+)}:\; ,  \label{bosonPsi}
\end{eqnarray}
and 
\begin{eqnarray}
&&\Psi^\dagger_F(x,t)=\rho(x)^{1/2} \sum_{m=-\infty}^{\infty} e^{iO_m}
:e^{i(\lambda^{1/2} + (2m+1)/\lambda^{1/2})X_{R}(x_-)}:  \nonumber \\
&&:e^{i(\lambda^{1/2}-(2m+1)/\lambda^{1/2})X_{L}(x_+)}:\; ,
\label{fermionPsi}
\end{eqnarray}
where the hermitian, constant-valued operators $O_m$ satisfy $[O_m,
O_{m^{\prime}}]=i\pi(m-m^{\prime})$, which give rise to the Klein factor
necessary for the correct commutation relations for $\Psi_{B(F)}$ and $%
\Psi_{B(F)}^{\dagger}$.

It is clear that the correlation functions of the bosons are different from
those of the fermions, e.g., for $m=0$, one has respectively 
\begin{equation}
\langle\Psi_B(x)\Psi^\dagger_B(0)\rangle \sim x^{-\lambda/2},
\end{equation}
and 
\begin{equation}
\langle\Psi_F(x)\Psi^\dagger_F(0)\rangle \sim x^{-(\lambda+1/\lambda)/2}\cos
k_F x .
\end{equation}

In summary, we have bosonized the C-S model and shown that the low energy
behavior is controlled by $c=1$ CFT. Depending on whether the C-S
interactions refer to those between boson or between fermions, we have two
different classes of $c=1$ CFT, governed by different selection rules for
quasi-particle quantum numbers. To our knowledge, the appearance of {\it two
different} classes of $c=1$ CFT's in the low energy limit of the C-S model
was {\it not} notified before in the literature.

In addition to the above charge-1 operators, there are more allowed
operators, the so-called anyons or exclusons, in the theory constructed as
follows: 
\begin{equation}
\Psi^\dagger_{B,\lambda}(x)= :\Psi^\dagger_{B}(x)e^{i\lambda^{1/2}
(X_R(x)-X_L(x))},
\end{equation}
for boson case and 
\begin{equation}
\Psi^\dagger_{F,\lambda}(x)= :\Psi^\dagger_{F}(x)e^{i(\lambda^{1/2}
-1/\lambda^{1/2})(X_R(x)-X_L(x))},
\end{equation}
for fermion case. Those operators satisfy 
\[
\Psi^\dagger_{\lambda}(x) \Psi^\dagger_{\lambda}(x^{\prime})- e^{i\pi\lambda
sgn(x-x^{\prime})}\Psi^\dagger_{\lambda}(x^{\prime})
\Psi^\dagger_{\lambda}(x)=0 
\]
for $x\not{= }x^{\prime}.$ The multi-sector density operator for exclusons
is 
\begin{eqnarray}
&&\hat{\rho}(x)=\Psi^\dagger_{\lambda}(x)
\Psi_{\lambda}(x)=\Psi_{B(F)}^\dagger(x) \Psi_{B(F)}(x)  \nonumber \\
&=& \rho(x)\sum{}_m :\exp\{i2m[X_R(x)-X_L(x)]/ \lambda^{1/2}\}:  \label{rhoh}
\end{eqnarray}

The correlation functions for density fluctuations and for anyons (or
exclusons) are the same in the above-mentioned two classes of $c=1$CFT's: 
\begin{eqnarray}
\langle \hat{\rho}(x,t)\hat{\rho}(0,0)\rangle &\approx&\bar{d}_0^2 \Biggl[1+%
\displaystyle \frac{1}{(2\pi\bar{d}_0)^2\lambda}\Biggl( \frac{1}{x_+^2}+%
\frac{1}{x_-^2}\Biggr)  \nonumber \\
&+&{\displaystyle\sum_{m=1}^{\infty}} A_m \frac{1}{[x_+x_-]^{m^2/\lambda}}
\cos(2\pi\bar{d}_0mx)\Biggr],
\end{eqnarray}
\begin{eqnarray}
G(x,t;\lambda)&\equiv& \langle \Psi^\dagger_{\lambda}(x,t)
\Psi_{\lambda}(0,0)\rangle  \nonumber \\
\approx \bar{d}_0\displaystyle \sum_{m=-\infty}^{\infty} &B_m&\frac{1}{%
x_-^{(m+\lambda)^2/\lambda}} \frac{1}{x_+^{m^2/\lambda}} e^{i(2\pi(m+%
\lambda/2)x+\mu t)},  \label{green}
\end{eqnarray}
with $A_m$ and $B_m$ regularization-dependent constants. We notice that the
correlation functions (\ref{green}) coincide with the asymptotic ones
calculated in \cite{Ha} for the C-S model.

\section{The Lieb-Wu solution and relevant developments for Hubbard model}

In the previous section, we achieved bosonization of the C-S model which has
no internal degree of freedom. The bosonization of 1-d models with internal
degrees of freedom such as spin is more complicated, because there are
backward scatterings and umklapp scatterings. We will take the Hubbard model
as an example to investigate. In this section, we first review Lieb-Wu's
Bethe ansatz solution and other developments which are relevant in the
coming sections.

The general form of the Hubbard Hamiltonian reads 
\begin{equation}
H=\displaystyle-t\sum_{\langle ij\rangle ,\sigma }c_{i\sigma }^{\dagger
}c_{j\sigma }+U\sum_in_{i\uparrow }n_{i\downarrow }+\mu \sum_in_i%
\displaystyle-\frac h2\sum_i(n_{i\uparrow }-n_{i\downarrow }),  \label{HubbH}
\end{equation}
where the summation $\langle ij\rangle $ runs over nearest-neighbors. We
have set the electrons on a 1-d ring with size $L$. This model has been
exactly solved by Lieb and Wu \cite{LbWu}, and the spectrum of the model is
shown to be determined by the following Bethe ansatz equations, 
\begin{eqnarray}
\displaystyle Lk_i &=&2\pi I_i+\sum_\alpha ^{N_{\downarrow }}2\tan ^{-1}%
\frac{4(\sin k_i-\lambda _\alpha )}u,  \nonumber \\
\displaystyle \sum_{i=1}^N &2&\tan ^{-1}\frac{4(\lambda _\alpha -\sin k_i)}u%
=2\pi J_\alpha +\sum_{\beta =1}^{N_{\downarrow }}2\tan ^{-1}\frac{2(\lambda
_\alpha -\lambda _\beta )}u,  \label{BetheA}
\end{eqnarray}
where $N$ ($N_{\downarrow }$) is the number of electrons (with down-spin)
and $u=U/t$. The total energy and momentum are related to the pseudomomenta $%
k_i$ by 
\begin{eqnarray}
E &=&\displaystyle -2t\sum_{i=1}^N\cos k_i+\mu N+h(N_{\downarrow }-\frac N2),
\nonumber \\
P &=&\displaystyle\sum_{i=1}^Nk_i=\frac{2\pi }L\Biggl(\sum_iI_i+\sum_\alpha
J_\alpha \Biggr).  \label{TEM}
\end{eqnarray}
Although the variation of rapidity $\lambda _\alpha $ seems to be related to
only the density variation of spin-down electrons in the $\lambda $-space,
it reflects the {\it spinon} density fluctuation in the {\it real} space,
because of the antiferromagnetic ground state. Hence, the {\it spin wave
rapidity} $\lambda _\alpha $ characterizes the spinon dynamics of the
system. The quantum numbers $I_i$ and $J_\alpha $ obey the selection rules 
\begin{equation}
I_i=\frac{N_{\downarrow }}2{\rm mod}(1),\hspace{0.2in}J_\alpha =\frac{%
N-N_{\downarrow }+1}2{\rm mod}(1).  \label{selectH}
\end{equation}
We would like to discuss the low energy behavior of the model with {\it %
repulsive} interaction ($U>0$) at {\it less than half filling}. We have two
kinds of gap-less excitations. Fortunately, it has been proven that the
excitations with complex $k$ and $\lambda $ have gaps of the order $u=U/t$ 
\cite{Woy}. We can only consider the real $k$ and $\lambda $ in the low
energy limit. It has been shown that the finite-size corrections to the
thermodynamic potential at $T=0$ are \cite{W-FK} 
\begin{equation}
\Omega _L(0)-\Omega _\infty (0)=-\frac \pi {6L}(v_c+v_s)+O(1/L^2),
\label{FSL}
\end{equation}
which implies that the low energy effective theory of the Hubbard model is
not Lorentz invariant. There are two Fermi velocities $v_c$ and $v_s$, which
correspond to the charge- and spin-wave velocities respectively, and hence
four Fermi points $\pm k_F$ and $\pm \lambda _F$. The finite-size scaling (%
\ref{FSL}) also shows that the low energy spectrum can be characterized by
two $c=1$ Virasoro algebras. To derive (\ref{FSL}), it is important to
realize that not only the thermodynamic potential (or the total energy) is
scaled but also the distributions of the charge and spin-wave. In the next
section, we will give (\ref{FSL}) another consistent check from finite-size
scaling in the temperature direction.

Again, we proceed to consider the thermodynamic limit. The zero temperature
behavior has been well-understood recently \cite{W-FK}. Around the Fermi
points, there are three kinds of excitations. The first kind is charge- and
spin-wave fluctuations in real space, which corresponds to non-zero modes in 
$k$-$\lambda$ space. We shall leave those to the coming sections. Other two
kinds are corresponding to zero modes in $k$-$\lambda$ space. One of them is
created by adding extra number of particles to the ground state. The other
is persistent currents. Both of excitations shift the ground state energy.
It is known that the ground state energy per site is given by 
\begin{equation}
\varepsilon_0=\frac{1}{2\pi}\int_{-k_F}^{k_F} dk ~\varepsilon_c(k),
\label{zero}
\end{equation}
where $\varepsilon _c$ is the zero-temperature dressed energy obeying 
\begin{eqnarray}
\varepsilon_c(k)&=&\varepsilon_c^{(0)} +\frac{1}{2\pi}\int_{-\lambda_F}
^{\lambda_F} d\lambda K_1(\sin k-\lambda) \varepsilon_s(\lambda),  \nonumber
\\
\varepsilon_s(\lambda)&=&\varepsilon_s^{(0)} +\frac{1}{2\pi}%
\int_{-k_F}^{k_F} dk\cos k K_1(\lambda-\sin k)\varepsilon_c(k)  \nonumber \\
&&-\frac{1}{2\pi}\int_{-\lambda_F}^{\lambda_F}
d\lambda^{\prime}K_2(\lambda-\lambda^{\prime})
\varepsilon_s(\lambda^{\prime}).  \label{DEZ}
\end{eqnarray}
The bare energies $\varepsilon_c^{(0)}=\mu-h/2-2t\cos k$ And $%
\varepsilon_s^{(0)}=h$. And the functions $K_1$ and $K_2$ are defined by 
\begin{equation}
K_1(x)=\frac{8u}{u^2+16x^2},\hspace{.2in} K_2(x) =\frac{4u}{u^2+4x^2}.
\end{equation}
The solutions of the integral equations (\ref{DEZ}) define the energy bands.
So, we can determine the $k_F$ and $\lambda_F$ by 
\begin{equation}
\varepsilon_c(\pm k_F)=0,\hspace{.2in} \varepsilon_s(\pm \lambda_F)=0.
\label{Z}
\end{equation}

Alternatively, we can fix the values of $k_F$ and $\lambda_F$ through the
particle numbers: 
\begin{eqnarray}
n_c&=&\frac{N}{L}=\int_{-k_F}^{k_F}dk \rho_c(k),  \nonumber \\
n_\downarrow&=&\frac{N_\downarrow}{L} =\int_{-\lambda_F}^{\lambda_F}d\lambda
\rho_\downarrow(\lambda),  \label{PNC}
\end{eqnarray}
where the distributions of charge- and spin-wave, $\rho(k)$ and $%
\rho_\downarrow(\lambda)$, are given by a set of integral equations: 
\begin{eqnarray}
\rho_c(k)&=&\frac{1}{2\pi}+\frac{\cos k}{2\pi} \int_{-\lambda_F}^{\lambda_F}
d\lambda K_1(\sin k-\lambda) \rho_\downarrow(\lambda),  \nonumber \\
\rho_\downarrow(\lambda)&=&\frac{1}{2\pi} \int_{-k_F}^{k_F} dk
K_1(\lambda-\sin k)\rho_c(k)  \nonumber \\
&-& \frac{1}{2\pi}\int_{-\lambda_F}^{\lambda_F}d\lambda^{\prime}K_2(\lambda-%
\lambda^{\prime}) \rho_\downarrow(\lambda^{\prime}).  \label{rho0}
\end{eqnarray}
Generally speaking, the zero mode excitations can be thought as the
fluctuations of the Fermi surface. The corresponding shift to the ground
state energy is given by 
\begin{equation}
\delta E_0=\frac{\partial^2 E_0}{\partial k_F^2} (\delta k_F)^2+ \frac{%
\partial^2 E_0}{\partial \lambda_F^2}(\delta \lambda_F)^2,  \label{DE}
\end{equation}
where the fact (\ref{Z}) insures there are no cross derivatives. It is easy
to see $\partial^2E_0/\partial k_F^2\sim v_c$ and $\partial^2E_0/\partial
\lambda_F^2\sim v_s$, where 
\begin{equation}
v_c=\frac{\varepsilon^{\prime}_c(k_F)}{2\pi \rho_c(k_F)}, \hspace{.2in} v_s =%
\frac{\varepsilon^{\prime}_s(\lambda_F)} {2\pi \rho_\downarrow (\lambda_F)}.
\label{FV}
\end{equation}
are the charge- and spin-wave velocities.

Adding extra particles $M$ and $M_\downarrow$ to the ground state leads to a
deformation of the Fermi surface given by 
\begin{equation}
\begin{array}{ccccc}
\left( 
\begin{array}{c}
\delta_Mk_F \\ 
\delta_M\lambda_F
\end{array}
\right) & = & \left( 
\begin{array}{cc}
\displaystyle \frac{\partial k_F}{\partial N} & \displaystyle \frac{\partial
k_F}{\partial N_\downarrow} \\ 
\displaystyle \frac{\partial \lambda_F}{\partial N} & \displaystyle \frac{%
\partial \lambda_F}{\partial N_\downarrow}
\end{array}
\right) & \left( 
\begin{array}{c}
M \\ 
M_\downarrow
\end{array}
\right) & 
\end{array}
.  \label{MTM}
\end{equation}

For the persistent currents, which correspond to the quantum number shifts $%
I\to I+D_I$ and $J\to J+D_J$, we have the Fermi surface deformation as
follows 
\begin{equation}
\begin{array}{ccccc}
\left( 
\begin{array}{c}
\delta_Jk_F \\ 
\delta_J\lambda_F
\end{array}
\right) & = & \left( 
\begin{array}{cc}
\displaystyle \frac{\partial k_F}{\partial D_I} & \displaystyle\frac{%
\partial k_F}{\partial D_J} \\ 
\displaystyle\frac{\partial \lambda_F}{\partial D_I} & \displaystyle\frac{%
\partial \lambda_F}{\partial D_J}
\end{array}
\right) & \left( 
\begin{array}{c}
D_I \\ 
D_J
\end{array}
\right) & 
\end{array}
.  \label{MTJ}
\end{equation}

In general, the matrices in (\ref{MTM}) and (\ref{MTJ}) are related to the
dressed charge matrix of the theory \cite{W-FK}. For the case of zero
magnetic field, $h=0$, which we are interested in, we have 
\begin{eqnarray}
\displaystyle \delta_Mk_F=\frac{1}{2\xi}\frac{M_c}{L},~~~&\displaystyle %
\delta_M\lambda_F=\frac{1} {2\sqrt{2}}\frac{M_s}{L},  \nonumber \\
\delta_Jk_F=\frac{\xi}{2}\frac{J_c}{L},~~~&\displaystyle \delta_J\lambda_F=%
\frac{1}{\sqrt{2}}\frac{J_s}{L},  \label{DFF}
\end{eqnarray}
where $M_c=M$, $M_s=M_\uparrow-M_\downarrow$, $J_c=2D_I+D_J$ and $J_s=-D_J$.
And $\xi=\xi(\sin k_F)$ with the function $\xi(x)$ determined by integral
equation 
\begin{eqnarray}
\xi(x)&=& 1+\displaystyle\frac{1}{2\pi} \int_{-\sin k_F}^{\sin k_F}
K(x-x^{\prime})\xi(x^{\prime}),  \nonumber \\
\displaystyle K(x)&=&\int_{-\infty}^{\infty} \frac{e^{-|\omega|u/4+i\omega x}%
} {2\cosh \omega u/4}d\omega.
\end{eqnarray}
In particular, we have 
\begin{equation}
\rho_c(k)= \frac{1}{2\pi}+\frac{\cos k}{2\pi} \int_{-k_F}^{k_F}dk^{\prime}K(%
\sin k-\sin k^{\prime}) \rho_c(k^{\prime}),  \label{rhoZM}
\end{equation}
\begin{equation}
\varepsilon_c(k)=\varepsilon^{(0)}(k) +\frac{1}{2\pi}\int_{-k_F}^{k_F}dk^{%
\prime}\cos k^{\prime}K(\sin k-\sin k^{\prime}) \varepsilon_c(k^{\prime}),
\label{DEZM}
\end{equation}

Hence, the zero mode part of the excitation spectrum at $h=0$, according to
our above discussion, is 
\begin{equation}
\delta E_0=(\pi/2L)(v_{c,N} M_c^2+v_{c,J} J_c^2) +(\pi/2L)(v_{s,N} M_s^2
+v_{s,J} J_s^2),  \label{DEMJ}
\end{equation}
where the velocity relations are given by 
\begin{equation}
v_{c(s),N}=v_{c(s)}\lambda_{c(s)}, ~v_{c(s),J}=v_{c(s)}/\lambda_{c(s)},
\label{HVR}
\end{equation}
with $\lambda_c=\xi^{-2}$ and $\lambda_s=1/2$. The energy shift (\ref{DEMJ})
and the velocity relations (\ref{HVR}), for each component (spin or charge),
resemble those in the C-S model, (\ref{CSDE}) and (\ref{Velo}). Also, it is
easy to show the momentum shift caused by zero modes is given by 
\begin{equation}
\delta P_0=\pi(n_{0c}+M_c/L )J_c+\pi M_sJ_s/L,
\end{equation}
which can be thought of as the two-component generalization of the momentum
shift in (\ref{CSDE}).

The selection rules (\ref{selectH}) imply that the excitation quantum
numbers obey the following selection rules \cite{W-FK} 
\begin{equation}
D_I=\frac{M_c-M_\downarrow}{2}{\rm mod}(1), \hspace{.2in} D_J= \frac{M}{2}%
{\rm mod}(1).  \label{selectCS}
\end{equation}
For $J_c$ and $J_s$, the fact $M_c=M_s{\rm mod}(1)$ and (\ref{selectCS})
imply that 
\begin{equation}
J_c=\frac{M_c}{2}{\rm mod}(1), ~~~J_s=\frac{M_s}{2}{\rm mod}(1).
\label{selectJCS}
\end{equation}

\section{Low temperature behavior and spin-charge separation}

In this section, we would like to present the thermodynamics of the Hubbard
model at low temperature and show spin-charge separation for arbitrary $U>0$%
. Also, we will confirm the finite-size scaling result (\ref{FSL}).

\bigskip

\subsection{Statistics and thermodynamics}

\bigskip

Recently an exclusion statistics description has been developed for the
Bethe ansatz soluble models \cite{WuBe,FuKa,HKKW}. For the Hubbard model,
the statistics matrix is given by 
\begin{eqnarray}
g^{cc}(k,k^{\prime})&=&\delta(k-k^{\prime}),  \nonumber \\
g^{c\downarrow}(k,\lambda)&=& -\frac{4}{\pi u}\frac{\cos k}{1+16(\sin k
-\lambda)^2/u^2},  \nonumber \\
g^{\downarrow \downarrow}(\lambda,\lambda^{\prime})
&=&\delta(\lambda-\lambda^{\prime})+\frac{1}{\pi} \frac{1}{%
1+4(\lambda-\lambda^{\prime})^2/u^2},  \nonumber \\
g^{\downarrow c}(\lambda,k)&=&-\frac{1}{\pi} \frac{2}{1+16(\lambda-\sin
k)^2/u^2}.  \label{stati}
\end{eqnarray}
The equation for $g^{\downarrow \downarrow}$ shows the mutual statistics
between different spin-wave rapidities, while $g^{c \downarrow}$ and $%
g^{\downarrow c}$ give the mutual statistics between $k$ and $\lambda$.
However, the constraint (\ref{Z}) means that the low energy excitations near
the Fermi surface have no mutual statistics between states described by $k$
and $\lambda$.

We showed before that for Bethe ansatz soluble models with no internal
degree of freedom, one can use an effective statistics parameter to
characterizes the statistics of low energy excitations, which can be read
off from the Luttinger-liquid velocity relations \cite{Wu}. For the present
case, the general velocity relation (\ref{HVR}) seems to suggest that there
is no effective mutual statistics and there are two effective statistics
parameters, $\lambda_c$ and $\lambda_s$, characterizing the statistics of
low energy excitations. Below, we will see that indeed there are two types
of anyonic (or exclusonic) excitations in the theory.

If only the real $k$ and $\lambda$ are taken into account, the thermodynamic
potential can be written down in terms of the principle of exclusion
statistics \cite{WuBe} and the statistics matrix (\ref{stati}). It has been
given in \cite{HKKW}: 
\begin{equation}
\frac{\Omega}{L}=-\frac{T}{2\pi} \int_{-\pi}^\pi dk \ln[1+e^{-%
\varepsilon_c(k,T)/T}],  \label{OmegaH}
\end{equation}
where the dressed energy $\varepsilon_c(k,T)$, which is the finite
temperature generalization of (\ref{DEZ}), is given by 
\begin{eqnarray}
\varepsilon_c(k,T)&=&\varepsilon_c^{(0)} -\frac{T}{2\pi}\int_{-\infty}^{%
\infty} d\lambda K_1(\sin k-\lambda)  \nonumber \\
&\times&\ln[1+e^{-\varepsilon_s(\lambda,T)/T}],  \nonumber \\
\varepsilon_s(\lambda,T)&=&\varepsilon_s^{(0)} -\frac{T}{2\pi}%
\int_{-\pi}^{\pi} dk \cos kK_1(\lambda-\sin k)  \nonumber \\
&\times&\ln[1+e^{-\varepsilon_c(k,T)/T}]  \nonumber \\
&&+\frac{T}{2\pi}\int_{-\infty}^{\infty}
d\lambda^{\prime}K_2(\lambda-\lambda^{\prime}) \ln[1+e^{-\varepsilon_s(%
\lambda^{\prime},T)/T}].  \label{DENZ}
\end{eqnarray}
The dressed energy $\varepsilon_s(\lambda,T)$ does not explicitly appear in (%
\ref{OmegaH}) since the number of bare single particle states for spinon is
zero.

Corresponding to the zero temperature dressed energy (\ref{DEZM}) at $h=0$,
its finite temperature counterpart is 
\begin{eqnarray}
&&\varepsilon_c(k,T)=\varepsilon_c^{(0)}  \nonumber \\
&&-\frac{T}{2\pi}\int_{-\pi}^{\pi}dk^{\prime}\cos k^{\prime}K(\sin k-\sin
k^{\prime})\ln[1+e^{-\varepsilon_c(k,T)}].  \label{Detzm}
\end{eqnarray}

Similarly, the charge- and spin-wave densities at finite temperature are
also given by a set of integral equations 
\begin{eqnarray}
&&\rho_c(k,T)=[1+e^{\varepsilon_c(k,T)/T}]^{-1} \biggl\{\frac{1}{2\pi} 
\nonumber \\
&&+\frac{\cos k}{2\pi} \int_{-\infty}^{\infty} d\lambda K_1(\sin k-\lambda)
\rho_\downarrow(\lambda,T)\biggr\},  \nonumber \\
&&\rho_\downarrow(\lambda,T) =[1+e^{\varepsilon_s(\lambda,T)/T}]^{-1} 
\nonumber \\
&& \biggl\{\frac{1}{2\pi}\int_{-\pi}^{\pi} dk K_1(\lambda-\sin k)\rho_c(k) 
\nonumber \\
&& -\frac{1}{2\pi}\int_{-\infty}^{\infty}d\lambda^{\prime}K_2(\lambda-%
\lambda^{\prime}) \rho_\downarrow(\lambda^{\prime}) \biggr\}.  \nonumber \\
&&  \label{rhoT}
\end{eqnarray}

\bigskip

\subsection{Finite-size scaling and spin-charge separation at low temperature
}

\bigskip

In this subsection, we would like to do finite-size scaling at finite
temperature, to show two new results. First, we want to confirm the
finite-size scaling (\ref{FSL}) at finite temperature, as promised above. On
the other hand, the spin-charge separation in the Hubbard model has been
previously shown only in the strong coupling limit for a given temperature 
\cite{OS,HKKW}. Here we want to show the spin-charge separation at low
temperature for arbitrary $U>0$, which was assumed in the Luttinger liquid
interpretation of the Hubbard model \cite{RA}.

The thermodynamic potential (\ref{OmegaH}) looks like that of a
single-component system (\ref{Omega}), because the dressed energy $%
\varepsilon _s(\lambda ,T)$ does not explicitly appear. However, it is
necessary to emphasize that $\varepsilon _c(k,T)$ is coupled to $\varepsilon
_s(\lambda ,T)$ through (\ref{DENZ}). Even at $T=0$, the dressed energies
are still coupled each other through (\ref{DEZ}). To see the finite-size
scaling in $T$, we fix $N=N_0$ and $N_{\downarrow }=N_0/2$. As shown in
Appendix B, $\varepsilon _{c(s)}(k(\lambda ),T)=\varepsilon
_{c(s)}(k(\lambda ),0)+O(\frac{T^2}\nu )$, i.e., one has 
\begin{eqnarray}
\varepsilon _c(k,T) &=&\varepsilon _c(k)+\tilde{\varepsilon}%
_c(k,T)+O(T^3/\nu ^2),  \nonumber \\
\varepsilon _s(\lambda ,T) &=&\varepsilon _s(\lambda )+\tilde{\varepsilon}%
_s(\lambda ,T)+O(T^3/\nu ^2).
\end{eqnarray}
According to (\ref{DENZ}), $\tilde{\varepsilon}_c(k,T)$ and $\tilde{%
\varepsilon}_s(\lambda ,T)$ can be determined by the following integral
equations 
\begin{eqnarray}
&&\tilde{\varepsilon}_c(k,T)=-\frac{\pi T^2}{6\varepsilon _c^{\prime }(k_F)}%
K_1(\sin k-\lambda _F)+\int_{-\lambda _F}^{\lambda _F}\frac{d\lambda }{2\pi }%
\tilde{\varepsilon}_s(\lambda ,T),  \nonumber \\
&&\tilde{\varepsilon}_s(\lambda ,T)=-\frac{\pi T^2}{6\varepsilon _s^{\prime
}(\lambda _F)}K_2(\lambda -\lambda _F)  \nonumber \\
&&-\frac{\pi T^2}{6\varepsilon _c^{\prime }(k_F)}K_1(\sin k_F-\lambda )\cos
k_F  \nonumber \\
&&\int_{-k_F}^{k_F}\frac{dk}{2\pi }\cos kK_1(\sin k-\lambda )\tilde{%
\varepsilon}_c(k,T)  \nonumber \\
&&-\int_{\lambda _F^{-}}^{\lambda _F^{+}}\frac{d\lambda ^{\prime }}{2\pi }%
K_2(\lambda -\lambda ^{\prime })\tilde{\varepsilon}_s(\lambda ,T).
\end{eqnarray}
Solving the integral equations, one finds that 
\begin{equation}
\int_{-k_F}^{k_F}\frac{dk}{2\pi }\tilde{\varepsilon}_c(k,T)=-\frac{\pi T^2}{%
6\varepsilon _c(k_F)}\tilde{f}_c-\frac{\pi T^2}{6\varepsilon _s(\lambda _F)}%
\tilde{f}_s,
\end{equation}
where 
\begin{equation}
\tilde{f}_c=2\pi \rho _c(k_F)-1,\hspace{0.2in}\tilde{f}_s=2\pi \rho
_s(\lambda _F).
\end{equation}
So, we have 
\begin{equation}
\frac 1{2\pi }\int_{-k_F}^{k_F}dk\tilde{\varepsilon}_c(k,T)=-\frac{\pi T^2}{%
6v_c}\biggl(1-\frac 1{2\pi \rho (k_F)}\biggr) -\frac{\pi T^2}{6v_s}.
\label{B1}
\end{equation}
Because we have fixed the particle numbers, the free energy can be expanded
as 
\begin{eqnarray}
\displaystyle &&\frac{F(T)}L=\frac{\Omega (T)}L-\mu N_0-\frac h2(N_{\uparrow
}-N_{\downarrow })  \nonumber \\
&\approx &\frac T{2\pi }\int_{-\pi }^\pi dk\ln [1+e^{-\varepsilon _c(k)/T}]+%
\frac T{2\pi }\int_{-\pi }^\pi \frac{dk}{1+e^{\varepsilon _c(k)/T}}\tilde{%
\varepsilon}_c(k,T)  \nonumber \\
&&-\mu N_0-\frac h2(N_{\uparrow }-N_{\downarrow })  \nonumber \\
&\approx &\frac{F(0)}L-\frac{\pi T^2}{6\varepsilon _c^{\prime }(k_F)}+\frac 1%
{2\pi }\int_{-k_F}^{k_F}dk\tilde{\varepsilon}_c(k,T),
\end{eqnarray}
where $F(0)/L\equiv \varepsilon _0$ ((\ref{zero})). Using (\ref{B1}), we
have 
\begin{equation}
\frac{F(T)}L-\frac{F(0)}L=-\frac{\pi T^2}{6v_c}-\frac{\pi T^2}{6v_s}.
\label{BM}
\end{equation}
In fact, this result is valid for arbitrary $N=N_0+M$ and $N_{\downarrow
}=N_0/2+M_{\downarrow }$ if $M\ll N_0$ and $M_{\downarrow }\ll N_0/2$.

Finally, we get the free energy in the low temperature limit, consistent
with (\ref{FSL}) and resulting in two $c=1$ Virasoro algebras. Combining (%
\ref{BM}) and (\ref{DEMJ}), we see that at $h=0$, the contributions of
excitations, including both zero modes and non-zero modes, to the
thermodynamic potential are separated into spin and charge parts. In this
sense, we can say that there is a spin-charge separation in the Hubbard
model at low temperature.

In the low-$T$ limit, the thermodynamic potential can be rewritten as 
\begin{eqnarray}
&&\frac{\Omega(T)}{L}-\frac{\Omega(0)}{L} =-\frac{2T}{\pi}%
I_s(\lambda_F,T)2\pi \rho_s(\lambda_F)  \nonumber \\
&&-\frac{2T}{\pi}I_c(k_F,T) 2\pi\rho_c(k_F),
\end{eqnarray}
where 
\begin{eqnarray}
I_s(\lambda_F,T)&=&\int_{\lambda_F}^{\lambda_F +\delta}d\lambda
\ln(1+e^{-\varepsilon_s(\lambda)/T}),  \nonumber \\
I_c(k_F,T)&=&\int_{k_F}^{k_F+\delta}dk \ln(1+e^{-\varepsilon_c(k)/T}.
\end{eqnarray}
The physical momenta corresponding to the charge and spin excitations are
defined by 
\begin{equation}
dp=2\pi\rho_c(k)dk,~~~d\tilde p =2\pi \rho_c(\lambda)d\lambda.
\end{equation}
Finally, the thermodynamic potential in the low-$T$ limit can be expressed
as 
\begin{eqnarray}
\frac{\Omega(T)}{L}&=&-\frac{T}{2\pi} \int_{-\pi}^\pi dk \ln[%
1+e^{-\varepsilon_c(k,\pm)/T)}]  \nonumber \\
&-&\frac{T}{\pi}\int_{-\delta}^\delta d\delta p_s
\ln(1+e^{-|\varepsilon_s(\delta p_s)|/T}),  \label{OmegaHE}
\end{eqnarray}
where $\delta p_s=p_s-p_s(\lambda_F)$. Eq. (\ref{OmegaHE}) is useful to
bosonize the Hubbard model in the next section.

\section{Bosonization of the Hubbard model}

Bosonization of the Hubbard model can generally be achieved by
Tomonaga-Luttinger's bosonization techniques \cite{ShSc}. However, the
existence of spin degree of freedom brings scattering processes other than
the forward scattering, say the backward and umklapp scatterings. They can
not be exactly diagonalized. Fortunately in the Hubbard model case, Luther
and Emery showed that the Hamiltonian still can be diagonalized at
particular values of coupling constants \cite{LuEm} and the backward and
umklapp scatterings can develop gaps for spin- and charge-waves
respectively. Then reasoning with renormalization group analysis \cite{Soly}%
, one expects that the backward and umklapp scatterings are irrelevant. In
this section, we will confirm these ideas in the bosonization approach based
on Bethe ansatz equations, which we have developed in Sec. II to bosonize
the C-S model.

\subsection{Bosonized Hamiltonian and Lagrangian}

Because of the similarity between (\ref{Omega}) and (\ref{OmegaH}), the
grand partition function of the Hubbard model can be put into a form like (%
\ref{PFCS}). And again this grand partition function corresponds to an ideal
fermion system with a complicated, $T$-dependent dressed energy $%
\varepsilon_c(k,T)$. To exploit this fermion representation, in the case of
the C-S model we noticed that $\epsilon(k,T)=\epsilon(k,0)
+O(e^{-|\epsilon|/T})$, so that we could use $\epsilon(k,0)$ to define an
effective Hamiltonian to derive the low-$T$ grand partition function.
However, this trick can not repeated for the Hubbard model, since $%
\varepsilon_c(k,T) =\varepsilon_c(k)+O(T^2/\nu)$. So we can not simply
ignore the $T$-dependent part of $\varepsilon_c(k,T)$ in discussing the low-$%
T$ thermodynamics of the system. 
%Hence, a similar effective Hamiltonian with 
%only one kind of fermions is not available. 

Instead of the original thermodynamic potential (\ref{OmegaH}), we begin
with the low-$T$ thermodynamic potential (\ref{OmegaHE}). Now, both $%
\varepsilon_c(k)$ and $\varepsilon_s(\delta{p}_s)$ are $T$-independent. In
principle, one may try to introduce two kinds of fermions with dispersions $%
\varepsilon_c(k)$ and $\varepsilon_s(\delta{p}_s)$ respectively. However,
the fermions with $|{p}_s|>{p}_{sF}+\delta$ and $|p_s|<p_{sF}-\delta$ are
not defined. (One may take these states to have zero energy, but it leads to
degeneracy deep inside the Fermi sea.) Rather, noting the cut-off $\delta$%
-independence of the low-$T$ thermodynamic potential, we first keep a finite 
$\delta$, and take the limit $\delta\to 0$ after all calculations. Thus we
rewrite (\ref{OmegaHE}) as 
\begin{eqnarray}
\frac{\Omega(T)}{L}&=&\lim_{\delta\to 0} \frac{\Omega(T,\delta)}{L}, 
\nonumber \\
\frac{\Omega(T,\delta)}{L}&=& \frac{\Omega_c(T)}{L}+\frac{\Omega_s(T,\delta)%
}{L},
\end{eqnarray}
with 
\begin{eqnarray}
\frac{\Omega_c(T)}{L}&=&-\frac{T}{2\pi} \int^\pi_{-\pi}dk
\ln(1+e^{-\varepsilon_c(k,\pm)/T}),  \nonumber \\
\frac{\Omega_s(T,\delta)}{L}&=& -\frac{T}{2\pi}\int^{n_\downarrow\pi}
_{-n_\downarrow\pi}d{p}_s \ln(1+e^{-{\varepsilon}_s(p_s,\delta)/T}).
\end{eqnarray}
Here ${\varepsilon}_s(p_s,\delta)$ is defined by 
\begin{equation}
{\varepsilon}_s(p_s,\delta) =\Biggl\{ {\ 
\begin{array}{lll}
\pm v_s(p_s\mp p_{sF}), \;\;\; & {p}_{sF}-\delta<|p_s|<p_{sF}+\delta, &  \\ 
\pm v_s (k\mp k_F) e^{-|p_s\mp{p}_{sF}|/\delta}, & 0\leq|p_s|< p_{sF}-\delta~
&  \\ 
~~ & {\rm or}~~|p_s|>{p}_{sF}+\delta. & 
\end{array}
}
\end{equation}
If we keep $\delta$ until all calculations are finished, the degeneracy of
the states far from the Fermi surface will be removed. Now both $%
\varepsilon_c(k)$ and ${\varepsilon}_s(p_s,\delta)$ become $T$-independent.
Therefore, the low-$T$ grand partition function can be obtained by an
effective Hamiltonian that incorporates two kinds of fermions with
dispersions $\varepsilon_c(k)$ and ${\varepsilon}_s(p_s,\delta)$: 
\begin{eqnarray}
Z_G&=&\prod_k(1+e^{-\varepsilon_c(k)/T}) \prod_{p_s} (1+e^{-{\varepsilon}%
_s(p_s,\delta)/T})  \nonumber \\
&=&{\rm Tr}e^{-\beta H_{{\rm eff}}}.
\end{eqnarray}
The effective Hamiltonian $H_{{\rm eff}}$ is given by 
\begin{equation}
H_{{\rm eff}}=\sum_k \varepsilon_c(k)c^\dagger_kc_k +\sum_{p_s}{\varepsilon}%
_s ({p}_s,\delta)s^\dagger_{p_s}s_{ p_s}+\delta E_0,
\end{equation}
where $c^\dagger_k$ and $s^\dagger_{p_s}$ are spinless fermion creation
operators, which can be identified as the charge- and spin-excitations.
Possible zero-mode excitations are included in the last term, $\delta E_0$,
of the effective Hamiltonian.

Now we are at a position to bosonize the theory, as we have done for the C-S
model, except that there are two kinds of fermions. Because of the
continuity of $\varepsilon^{\prime}_c(k)$ and ${\varepsilon}%
^{\prime}_s(p_s,\delta)$ at Fermi surface, the bosonization is standard. At
low-$T$, only low energy excitations near the Fermi surface are relevant.
So, instead of the full expressions of the dispersions, the linear
dispersions are employed 
\begin{eqnarray}
\varepsilon^\pm_c(k)&=&\pm \varepsilon^{\prime}_c(k_F)(k\mp k_F) =\pm
v_c(p\mp p_F),  \nonumber \\
{\varepsilon}^\pm_s(p_s)&=& \pm v_s(p_s \mp p_{sF}),
\end{eqnarray}
and the effective Hamiltonian is separated into the right- and left- moving
parts 
\begin{eqnarray}
H_{{\rm eff}}&=&H_{c,+}+H_{c,-}+H_{s,+}+H_{s,-} +\delta E_0,  \nonumber \\
H_{c,\pm}&=&\sum_p\varepsilon^\pm_c(k) c^\dagger_p c_p,~~~H_{s,\pm}
=\sum_{p_s} {\varepsilon}^\pm_s(p_s) s^\dagger_{p_s} s_{p_s}.
\end{eqnarray}
By using the well-known bosonization technique, we have 
\begin{eqnarray}
\rho_{c,q}^{(+)}&=&\sum_{p\sim p_F} :c^\dagger_{p+q}c_p:,\hspace{.2in}
\rho_{c,q}^{(-)}=\sum_{p\sim -p_F} :c^\dagger_{p-q}c_p:,  \nonumber \\
\rho_{s,q}^{(+)}&=&\sum_{p_s\sim p_{sF}} :s^\dagger_{p_s+q} s_{p_s}:, 
\hspace{.2in} \rho_{s,q}^{(-)}=\sum_{p_s\sim -p_{sF}} :s^\dagger_{p_s-q}
s_{p_s}:.
\end{eqnarray}
They satisfy the following commutation relations: 
\begin{eqnarray}
&[&\rho_{c(s),q}^{(\pm)},\rho_{c(s),q^{\prime}}^{(\pm)\dagger}] =\frac{L}{%
2\pi}q \delta_{q,q^{\prime}},  \nonumber \\
&[&H_{c(s),\pm},\rho_{c(s),q}^{(\pm)}] = \pm v_{c(s)}\rho_{c,(s),q}^{(\pm)},
\label{CRSH}
\end{eqnarray}
which describe two 1-d free phonons with the sound velocities $v_c$ and $v_s$%
. At $h=0$, the bosonized Hamiltonian satisfying (\ref{CRSH}) is given by 
\begin{eqnarray}
H_B&=&\displaystyle v_c\sum_{q>0}(a^\dagger_qa_q +\tilde{a}_q^\dagger\tilde{a%
}_q)+\frac{1}{2} \frac{\pi}{L}(v_{c,N}M_c^2+v_{c,J}J^2_c)  \nonumber \\
&&\displaystyle +v_s\sum_{q>0}(b^\dagger_qb_q +\tilde{b}_q^\dagger\tilde{b}%
_q)+\frac{1}{2} \frac{\pi}{L}(v_{s,N}M_s^2+v_{s,J}J^2_s),  \nonumber \\
&&  \label{HHB}
\end{eqnarray}
where 
\begin{eqnarray}
a_q&=&\sqrt{2\pi/qL}\rho_{c,q}^{(+)},\hspace{.2in} \tilde{a}_q=\sqrt{2\pi/qL}%
\rho_{c,q}^{(-)\dagger},  \nonumber \\
b_q&=&\sqrt{2\pi/qL}\rho_{s,q}^{(+)},\hspace{.2in} \tilde{b}_q=\sqrt{2\pi/qL}%
\rho_{s,q}^{(-)\dagger}
\end{eqnarray}
which are normalized boson annihilation operators. The cut-off $\delta$%
-dependence only contributes a constant to (\ref{HHB}), which vanishes as $%
\delta\to 0$. So, we can take $\delta\to 0$ at this stage, which does not
affect any results that we will obtain below.

The bosonized total momentum operator may also be obtained by bosonizing the
fermionized $P=\sum_pc^\dagger_pc_p +\sum_{p_s}s^\dagger_{p_s} s_{p_s}$. It
reads 
\begin{eqnarray}
P&=&\displaystyle\sum_{q>0}q(a^\dagger_qa_q -\tilde{a}_q^\dagger\tilde{a}_q)
+\pi J_c(n_{c0}+M_c/L)  \nonumber \\
&&\displaystyle+\sum_{q>0}q(b^\dagger_qb_q -\tilde{b}_q^\dagger\tilde{b}%
_q)+\pi J_s M_s/L.
\end{eqnarray}
\bigskip

In the coordinate-space formulation, the charge-density field and
spin-density field are given by 
\begin{equation}
\rho_{c(s)}(x)=\rho_{c(s),R}(x)+\rho_{c(s),L}(x),
\end{equation}
with 
\begin{eqnarray}
\rho_{c,R}(x)&=&\frac{M_{c,R}}{L}+\displaystyle \sum_{q>0}\sqrt{\frac{q}{%
2\pi L\lambda_c}} (e^{iqx}a_q+e^{-iqx}a_q^\dagger),  \nonumber \\
\rho_{c,L}(x)&=&\frac{M_{c,L}}{L}+\displaystyle \sum_{q>0}\sqrt{\frac{q}{%
2\pi L\lambda_c}} (e^{-iqx}\tilde{a}_q+e^{iqx}\tilde{a}_q^\dagger), 
\nonumber \\
\rho_{s,R}(x)&=&\frac{M_{s,R}}{L}+\displaystyle \sum_{q>0}\sqrt{\frac{q}{%
2\pi L\lambda_s}} (e^{iqx}b_q+e^{-iqx}b_q^\dagger),  \nonumber \\
\rho_{s,L}(x)&=&\frac{M_{s,L}}{L}+\displaystyle \sum_{q>0}\sqrt{\frac{q}{%
2\pi L\lambda_s}} (e^{-iqx}\tilde{b}_q+e^{iqx}\tilde{b}_q^\dagger).
\end{eqnarray}
Here $M_{c(s)}=M_{c(s),R}+M_{c(s),L}$. The conjugate field $\phi_{c(s)}(x)$
of $\rho_{c(s)}(x)$, which obeys $[\phi_{c(s)}(x),\rho_{c(s)}(x^{\prime})]
=i\delta(x-x^{\prime})$, is $\phi_{c(s)}(x)=\phi_{c(s),R}(x)+%
\phi_{c(s),L}(x) $ with 
\begin{eqnarray}
\phi_{c,R}(x)&=&\phi_{c0,R}+\frac{\pi J_{c,R}x}{L}+ \displaystyle i\sum_{q>0}%
\sqrt{\frac{\pi\lambda_c}{2qL}} (e^{iqx}a_q-e^{-iqx}a_q^\dagger),  \nonumber
\\
\phi_{c,L}(x)&=&\phi_{c0,L}+\frac{\pi J_{c,L}x}{L}+ \displaystyle i\sum_{q>0}%
\sqrt{\frac{\pi\lambda_c}{2qL}} (e^{-iqx}\tilde{a}_q-e^{iqx}\tilde{a}%
_q^\dagger),  \nonumber \\
\phi_{s,R}(x)&=&\phi_{s0,R}+\frac{\pi J_{s,R}x}{L}+ \displaystyle i\sum_{q>0}%
\sqrt{\frac{\pi\lambda_s}{2qL}} (e^{iqx}b_q-e^{-iqx}b_q^\dagger),  \nonumber
\\
\phi_{s,L}(x)&=&\phi_{s0,L}+\frac{\pi J_{s,L}x}{L}+ \displaystyle i\sum_{q>0}%
\sqrt{\frac{\pi\lambda_s}{2qL}} (e^{-iqx}\tilde{b}_q-e^{iqx}\tilde{b}%
_q^\dagger),  \nonumber \\
&&
\end{eqnarray}
where $J_{c(s)}=J_{c(s),R}+J_{c(s),L}$. Here $M_{c(s)}$ and $J_{c(s)}$ are
operators with integer eigenvalues obeying the selection rule (\ref
{selectJCS}), and $\phi_{c0}=\phi_{c0,R}+\phi_{c0,L}$ and $\phi_{s0}=
\phi_{s0,R}+\phi_{s0,L}$ are angular variables conjugated to $M_c$ and $M_s$%
: $[\phi_{c(s)0},M_{c(s)}]=i$.

In the coordinate-space formulation, the bosonized Hamiltonian (\ref{HHB})
reads 
\begin{eqnarray}
&H&_B=H_{B,c}+H_{B,s},  \nonumber \\
&H&_{B,c}=\frac{v_c}{2\pi}\int_0^L dx [\Pi_c(x)^2+(\partial_xX_c(x))^2], 
\nonumber \\
&H&_{B,s}=\frac{v_s}{2\pi}\int_0^L dx [\Pi_s(x)^2+(\partial_xX_s(x))^2],
\end{eqnarray}
where $\Pi_{c(s)}(x)= \pi\lambda_{c(s)}^{1/2}\rho_{c(s)}(x)$ and $%
X_{c(s)}=\lambda_{c(s)}^{-1/2}\phi_{c(s)}(x)$. With the Heisenberg operators 
$A(x,t)=e^{iH_Bt}A(x)e^{-iH_Bt}$, the Lagrangian density is given by 
\begin{eqnarray}
&{\cal L}&={\cal L}_c+{\cal L}_s,  \label{LT} \\
&{\cal L}&_c=\frac{v_c}{2\pi}\partial_\alpha X_c(x,t)\partial^\alpha
X_c(x,t),  \label{LC} \\
&{\cal L}&_s=\frac{v_s}{2\pi}\partial_\alpha X_s(x,t)\partial^\alpha
X_s(x,t).  \label{LS}
\end{eqnarray}
The Lagrangians (\ref{LC}) and (\ref{LS}) give rise to two $c=1$ Virasoro
algebras. We see that there is no mixture between spin and charge sectors in
the total Lagrangian (\ref{LT}). Therefore, in charge-spin basis, the low
energy behavior of the system is characterized by the direct product of two
independent Virasoro algebras both with central charges $c=1$ \cite{comm3}.

Note the selection rule (\ref{selectJCS}); then one can take $%
D_{c(s)}=2J_{c(s)}=M_{c(s)}{\rm mod}(1)$. The zero-mode part in $H_{B,c(s)}$
becomes 
\begin{equation}
\frac{v_{c(s)}\pi}{2L}\biggl(\lambda_{c(s)}M^2_{c(s)} +\frac{1}{%
4\lambda_{c(s)}}D^2_{c(s)}\biggr).
\end{equation}
So each sector has the duality relation as in the usual $c=1$ CFT.

Changing the right-and-left-moving representation to the $\theta$-$\phi$
representation with $\theta=\phi_R-\phi_L$, which Haldane has used in
discussing the Luttinger liquid theory \cite{Hald5}, we see that our
bosonization theory precisely agrees with the Luttinger liquid theory for
the Hubbard model proposed by Ren and Anderson\cite{RA}. The selection rules
chosen in \cite{RA} were 
\begin{equation}
J_{\uparrow,\downarrow}= M_{\uparrow,\downarrow} {\rm mod}(2),
\label{selectEL}
\end{equation}
where 
\begin{equation}
J_c=\frac{J_\uparrow+J_\downarrow}{2}, \hspace{.2in} J_s=\frac{%
J_\uparrow-J_\downarrow}{2}.  \label{JEL}
\end{equation}
The selection rules (\ref{selectEL}) of $J_\uparrow$ and $J_\downarrow$ are
a restriction on the quantum numbers more constrained than that given by (%
\ref{selectJCS}). For example, the choice $M_c=M_s=1$ ,$J_c=1/2$ and $%
J_s=-1/2$, which equals to $M_\uparrow=1$, $M_\downarrow=0$ and $%
J_\uparrow=0 $ and $J_\downarrow=1$, is forbidden by (\ref{selectEL}) but is
allowed by (\ref{selectJCS}). This choice corresponds to a charge-1, spin-up
boson excitation near $k_F$ as we will see below. We believe that (\ref
{selectJCS}) gives the correct selection rules for the quantum numbers,
since it is derived on the basis of the Bethe ansatz equations. 
%it is in some sense a ` more exact' approximation 
%of the full theory in the low energy limit.

\subsection{Correlation functions of single quasiparticles}

In order to calculate the correlation functions in the low energy limit, we
first need to determine the {\it allowed} operator in the theory. Since
there are two angular variables, $\phi_{c0}$ and $\phi_{s0}$, there are
hidden symmetries in the theory under 
\begin{equation}
\phi_{c(s)}\to\phi_{c(s)}+2\pi.  \label{phi}
\end{equation}
Then $X_{c(s)}$ is compactified on a circle with radius $R_{c(s)}=%
\lambda^{-1/2}_{c(s)}$, 
\begin{equation}
X_{c(s)}\sim X_{c(s)}+2\pi R_{c(s)}.  \label{X}
\end{equation}
Operators $V[X]$ are {\it allowed} if only if they are invariant under (\ref
{phi}) (or (\ref{X})). Because the theory is described by the direct product
of two Virasoro algebras, the primary fields of the theory have a similar
structure. The primary fields satisfy the periodic boundary conditions are 
\begin{eqnarray}
\phi_{[M_c,J_c;M_s,J_s]}(x)&\sim& :
e^{i(M_c\lambda_c^{1/2}+J/\lambda_c^{1/2})X_{R,c}(x)}  \nonumber \\
&& e^{i(M_c\lambda_c^{1/2}-J_c/\lambda_c^{1/2})X_{L,c}(x)}  \nonumber \\
&&e^{i(M_s\lambda_s^{1/2}+J_s/\lambda_s^{1/2})X_{R,s}(x)}  \nonumber \\
&&e^{i(M_s\lambda_s^{1/2}-J_s/\lambda_s^{1/2})X_{L,s}(x)}:.
\end{eqnarray}
The fermion operator, say of charge-1 and spin-up, is given by $M_c=M_s=1$: 
\begin{eqnarray}
\Psi^\dagger_F(x)&\sim& \sum_{J_c+J_s ={\rm odd}}
e^{iO_J}:e^{i(\lambda_c^{1/2} +J/\lambda_c^{1/2})X_{R,c}(x)}  \nonumber \\
&& e^{i(\lambda_c^{1/2} -J_c/\lambda_c^{1/2})X_{L,c}(x)}  \nonumber \\
&&e^{i(\lambda_s^{1/2} +J_s/\lambda_s^{1/2})X_{R,s}(x)}  \nonumber \\
&&e^{i(\lambda_s^{1/2} -J_s/\lambda_s^{1/2})X_{L,s}(x)}:,  \label{FN}
\end{eqnarray}
where $O_J$ are constant value operators satisfying $[O_J,O_{J^{%
\prime}}]=i[(J_c+J_s)/2-(J_c^{\prime}+J_s^{\prime})/2]$.

The above fermion operator satisfies our selection rule (\ref{selectJCS})
and, since $J_c+J_s={\rm odd}$, the selection rule (\ref{selectEL}) of Ren
and Anderson as well. However, our rule (\ref{selectJCS}) allows more
possible quantum numbers, say charge-1, spin-up bosonic excitations
described by the primary fields: 
\begin{eqnarray}
\Psi^\dagger_B(x)&\sim& \sum_{J_c+J_s={\rm even}} e^{iO_J}
:e^{i(\lambda_c^{1/2} +J/\lambda_c^{1/2})X_{R,c}(x)}  \nonumber \\
&& e^{i(\lambda_c^{1/2} -J_c/\lambda_c^{1/2})X_{L,c}(x)}  \nonumber \\
&&e^{i(\lambda_s^{1/2} +J_s/\lambda_s^{1/2})X_{R,s}(x)}  \nonumber \\
&&e^{i(\lambda_s^{1/2}-J_s/\lambda_s^{1/2})X_{L,s}(x)}:.  \label{BN}
\end{eqnarray}
The selection rule for bosons reads 
\begin{equation}
J_{\uparrow,\downarrow} =(M_{\uparrow,\downarrow}+1){\rm mod}(2),
\label{selectB}
\end{equation}
which is different with that for fermions, (\ref{selectEL}).

Besides the fermion and boson excitations, there are two kinds of anyonic
excitations which are charge-1 but non-periodic primary fields. The anyon
operators read 
\begin{eqnarray}
\Psi^\dagger_{\lambda_c}=\Psi^\dagger_B
e^{i\lambda^{1/2}_c(X_{R,c}-X_{L,c})},  \nonumber \\
\Psi^\dagger_{\lambda_s}=\Psi^\dagger_B
e^{i\lambda^{1/2}_s(X_{R,s}-X_{L,s})},  \label{AN}
\end{eqnarray}
which obey 
\begin{equation}
\Psi^\dagger_{\lambda_{c(s)}}(x) \Psi^\dagger_{\lambda_{c(s)}}(x^{\prime})
-e^{i\lambda_{c(s)}\pi}\Psi^\dagger_{\lambda_{c(s)}}(x^{\prime})
\Psi^\dagger_{\lambda_{c(s)}}(x)=0.
\end{equation}
Thus, the statistics of $\Psi^\dagger_{\lambda_{c}}$ is
interaction-dependent, while $\Psi^\dagger_{\lambda_{s}}$ is a semion, in
the half way between fermion and boson.

The dynamic correlation functions of fermionic and bosonic single-particle
operators, (\ref{FN}) and (\ref{BN}), can be easily calculated: 
\begin{eqnarray}
&&G_F(x,t)\equiv \langle\Psi_F(x,t) \Psi^\dagger_F(0,0)\rangle  \nonumber \\
&&\sim\sum_{n=-\infty,m=4n+1}^{\infty}A_m  \nonumber \\
&& x_{R,c}^{-\frac{1}{8}(e^{-\varphi}+m e^{\varphi})^2}x_{L,c}^{-\frac{1}{8}
(e^{-\varphi}-me^{\varphi})^2}x_{R,s}^{-\frac{1}{2}} e^{imk_F}  \nonumber \\
&&+\sum_{n=-\infty,m=4n+3}^{\infty}B_m  \nonumber \\
&&x_{R,c}^{-\frac{1}{8}(e^{-\varphi}+m e^{\varphi})^2}x_{L,c}^{-\frac{1}{8}
(e^{-\varphi}-me^{\varphi})^2}x_{L,s}^{-\frac{1}{2}} e^{imk_F},  \label{CFF}
\end{eqnarray}
\begin{eqnarray}
&&G_B(x,t)\equiv \langle\Psi_B(x,t) \Psi^\dagger_B(0,0)\rangle  \nonumber \\
&&\sim\sum_{n=-\infty,m=4n+1}^{\infty}C_m  \nonumber \\
&& x_{R,c}^{-\frac{1}{8}(e^{-\varphi}+m e^{\varphi})^2}x_{L,c}^{-\frac{1}{8}
(e^{-\varphi}-me^{\varphi})^2} x_{L,s}^{-\frac{1}{2}} e^{imk_F}  \nonumber \\
&&+\sum_{n=-\infty,m=4n+3}^{\infty}D_m  \nonumber \\
&&x_{R,c}^{-\frac{1}{8}(e^{-\varphi}+m e^{\varphi})^2}x_{L,c}^{-\frac{1}{8}
(e^{-\varphi}-me^{\varphi})^2} x_{R,s}^{-\frac{1}{2}} e^{imk_F},  \label{CFB}
\end{eqnarray}
where $A_m$, $B_m$, $C_m$ and $D_m$ are regularization-dependent constants,
and $x_{R,c(s)}=x-v_{c(s)}t$ and $x_{L,c(s)}=x+v_{c(s)}t$. $%
e^{-2\pi}=2\lambda_c$. To derive (\ref{CFF}) and (\ref{CFB}), we have taken
the leading term only for each given $m$. For the fermion case, $J_s$ is
taken to be $\frac{1}{2}$ or $-\frac{1}{2}$ with respect to $m=4n+1$ or $%
4n+3 $. For the boson case, $J_s$ is taken to be $-\frac{1}{2}$ or $\frac{1}{%
2}$ with respect to $m=4n+1$ or $4n+3$. We see that the only difference
between fermion and boson is that when the fermion has a left(right)-moving
spinon-part in a given $m k_F$-oscillation, the boson has a
right(left)-moving one. They have the same momentum distribution
singularities: 
\begin{equation}
n(k)\sim {\rm const}.-{\rm const.~sgn}(k-k_F) |k-k_F|^{(e^{-2\varphi}+
e^{2\varphi}-2)/4},
\end{equation}
for $k_F$-oscillations and 
\begin{equation}
n(k)\sim {\rm const}.-{\rm const.~sgn}(k-3k_F) |k-3k_F|^{(9e^{-2\varphi}+
e^{2\varphi}-2)/4 },
\end{equation}
for $3k_F$-oscillations, etc. They are consistent with those given in \cite
{RA}. In the strong coupling limit, $U/t\to \infty$, one has $%
(e^{-2\varphi}+e^{2\varphi}-2)/4=\frac{1}{8}$ and $(9e^{-2\varphi}+e^{2%
\varphi}-2)/4=\frac{9}{8}$ as $e^{-2\varphi}=\frac{1}{2}$. This implies that
there is a similarity of boson and fermion in the two-component Luttinger
liquid theory.

The anyon correlation functions read 
\begin{eqnarray}
&&G_{\lambda_c}(x,t)\equiv \langle\Psi_{\lambda_c}(x,t)
\Psi^\dagger_{\lambda_c}(0,0)\rangle  \nonumber \\
&&\sim\sum_{n=-\infty,m=4n+1}^{\infty} A^c_m x_{R,c}^{-\frac{1}{8}%
(2e^{-\varphi} +me^{\varphi})^2}  \nonumber \\
&&x_{L,c}^{-\frac{1}{8}(me^{\varphi})^2} x_{L,s}^{-\frac{1}{2}} e^{imk_F} 
\nonumber \\
&&+\sum_{n=-\infty,m=4n+3}^{\infty}B^c_m x_{R,c}^{-\frac{1}{8}%
(2e^{-\varphi}+m e^{\varphi})^2}  \nonumber \\
&&x_{L,c}^{-\frac{1}{8}(me^{\varphi})^2} x_{R,s}^{-\frac{1}{2}}e^{imk_F}, 
\nonumber \\
~~  \label{CRLC}
\end{eqnarray}
and 
\begin{eqnarray}
&&G_{\lambda_s}(x,t)\equiv \langle\Psi_{\lambda_s}(x,t)
\Psi^\dagger_{\lambda_s}(0,0)\rangle  \nonumber \\
&&\sim\sum_{n=-\infty,m=4n+1}^{\infty} A^s_m x_{R,c}^{-\frac{1}{8}%
(e^{-\varphi}+me^{\varphi})^2} x_{L,c}^{-\frac{1}{8}(e^{-\varphi}-me^{%
\varphi})^2} x_{R,s}^{-\frac{9}{8}} x_{L,s}^{-\frac{1}{8}}e^{imk_F} 
\nonumber \\
&&+\sum_{n=-\infty,m=4n+3}^{\infty}B^s_m x_{R,c}^{-\frac{1}{8}%
(e^{-\varphi}+m e^{\varphi})^2} x_{L,c}^{-\frac{1}{8}(e^{-\varphi}-m
e^{\varphi})^2} x_{R,s}^{-\frac{1}{8}}x_{L,s}^{-\frac{1}{8}}e^{imk_F}.
\label{CRLS}
\end{eqnarray}

The momentum distributions corresponding to (\ref{CRLC}) vanish faster than
those for fermions or bosons near $mk_F$. So, this kind of excitations may
be more difficult to be observed. The distributions corresponding to (\ref
{CRLS}) near $(4n+1)k_F$, say $k_F$, also decay fast but those near $%
(4n+3)k_F$, say $3k_F$, do decay slower: 
\begin{eqnarray}
n(k)&\sim&{\rm const}.-{\rm const.~sgn}(k-3k_F) |k-3k_F|^{(9e^{-2\varphi}+
e^{2\varphi}-3)/4}  \nonumber \\
&\to& {\rm const}.-{\rm const.~sgn}(k-3k_F) |k-3k_F|^{\frac{7}{8}%
},~~U/t\to\infty.
\end{eqnarray}
So, the semion may possibly be observed.

As shown by (\ref{rhoh}), for the C-S model, it is the same density operator
no matter that the single-particle excitations are fermionic, bosonic or
anyonic. This remains true for the charge density and spin density in the
Hubbard model. And so for the charge-charge and spin-spin density
correlation functions. We can define the pairing operators in the theory.
One finds that the most important contributions to the spin-singlet and
triplet pairings, up to $2k_F$-oscillation, still come from the fermionic
(or bosonic) excitations. All of these correlation functions have been
calculated in the reference \cite{RA}, and can be reproduced easily within
our approach.

\section{Conclusions}

We have developed a new approach for bosonizing 1-d exactly soluble
many-body models at low energy. In this approach, one first derives the low
temperature partition function from the Bethe ansatz equations, and then
bosonizes the partition function. The asymptotic correlation functions can
then be calculated by field theory techniques. Because our bosonization
started with the Bethe ansatz equations, we call it the bosonization based
on Bethe ansatz equations. We explicitly carried out this program with the
C-S model and the Hubbard model as examples. Some new results were obtained
during this exercise.

For the C-S model, we showed that the low energy effective theory is
described by two different classes of $c=1$ CFT's, depending on whether the
Hamiltonian describes the interactions between bosons or between fermions.
Using the standard terminology in CFT \cite{CFT}, it is a $c=1$ CFT
compactified on a circle $S^1$ for the bosonic case, compactified on an
interval $S^1/Z_2$ for the fermionic case. A piece of evidence for this
difference is that the charge-1 primary field in the two theories are
different. It is a fermion in the theory of interacting fermions, while it
is a boson for the interacting bosons. Their correlation functions have
totally different asymptotic behavior. Also our bosonization of the C-S
model prepares some techniques for bosonizing multi-component models.

The repulsive Hubbard model at less than half filling is an example of the
two-component models. In the above the spin-charge separation in this model
has been explicitly verified not only for the strong-coupling limit but also
for {\it finite} $U$. The low energy theory is shown to be controlled by the
direct product of two usual $c=1$ Virasoro algebras. We confirmed this point
by the finite-size scaling in the temperature direction. Furthermore, the
explicit low-$T$ grand partition function enabled us to identify the system
with a two-component Luttinger liquid theory by using the bosonization
technique. We showed that the backforward scattering processes are
suppressed at low energy, in agreement with the renormalization group
analysis. We pointed out that not only the fermionic excitations are allowed
near the Fermi surface, but also the bosonic and anyonic ones. The
single-particle correlation functions of those excitations were
systematically calculated in a simple way.

Our bosonization approach can be easily generalized to any 1-d many-body
models that is exactly soluble by Bethe ansatz, including thermodynamic
Bethe ansatz, and provides a simple way to calculate asymptotic correlation
functions using field theory techniques.

This work was supported in part by the US NSF grants PHY-9309458 PHY-9970701
and NSF of China.

\appendix

\section{Finite size scaling in the L-direction for the C-S model}

This appendix is devoted to resolve a puzzle raised in \cite{KaYa}, where it
was found that if we take $k=\frac{2\pi \lambda}{L}n$ in the thermodynamic
limit, naively the discrete sum for the total energy of the C-S model
differs from the continuous integration by $-\frac{\pi\lambda v_s}{6L}$.
This seems to imply a central charge $c=\lambda$, in conflict to (\ref{CE1}%
). However, as we see below, a careful consideration for the finite size
scaling shows that the naive result is wrong. There is an additional term of
order $O(1/L^3)$ when we go to the continuous limit of (\ref{ABA}) from the
discrete version. This will cause a rescaling of the particle density
together with that of the total energy. Taking into account this subtlety,
we will finally achieve the correct finite size scaling of the ground state
energy 
\begin{equation}
E_{0,L}-E_0=-\frac{\pi v_s}{6L}+O(\frac{1}{L^2}).
\end{equation}

To see this, a useful formula relating the discrete sum to the integration
is 
\begin{eqnarray}
\frac{1}{L}\sum_{n=N_1}^{N_2}f(\frac{I_n}{L})
&=&\int_{(N_1+1/2)/L}^{(N_2-1/2)/L}dx f(x)  \nonumber \\
&+&\frac{1}{24L^2}[f^{\prime}((N_1-1/2)/L)  \nonumber \\
&-& f^{\prime}((N_2+1/2)/L)]+O(1/L^3).  \label{FSS}
\end{eqnarray}
Using (\ref{FSS}), the discrete version of the density $\rho_L(k)$ can be
written as 
\begin{eqnarray}
\rho_L(k)&=&\frac{1}{2\pi}+(1-\lambda)\rho_L(k)  \nonumber \\
&+&\frac{\pi}{12L^2}(1-\lambda) \lambda\frac{d}{dk_F}[\delta(k-k_F)-%
\delta(k+k_F)].
\end{eqnarray}
The discrete version differs from its continuous counterpart by a term of
the order $O(1/L^2)$. Denote 
\begin{equation}
\rho_L(k)=\rho(k)+\rho_1(k);
\end{equation}
one has 
\begin{equation}
\rho_1(k)=-\frac{\pi}{12L^2}(1-\lambda) [\frac{d}{dk_F}\delta(k+k_F) -\frac{d%
}{dk_F}\delta(k-k_F)],
\end{equation}
while $\rho(k)=1/2\pi \lambda$. Now let's examine the finite size scaling of
the ground state energy. Using (\ref{FSS}) again, we have 
\begin{eqnarray}
\frac{E_L}{L}&=&\int_{-k_F}^{k_F}dk \rho_L(k)\epsilon_0  \nonumber \\
&+&\frac{1}{24L^2\rho(k_F)}[\epsilon^{\prime}_0(k) \biggl|%
_{-k_F}-\epsilon^{\prime}_0(k)\biggl|_{k_F}]  \nonumber \\
&=&\int_{-k_F}^{k_F}dk\rho(k)\epsilon_0(k)+
\int_{-k_F}^{k_F}dk\rho_1(k)\epsilon_0(k) -\frac{\pi \lambda v_F}{6L^2}.
\label{GSE1}
\end{eqnarray}
The second term of the last equation is easy to calculate and one has 
\begin{equation}
\int_{-k_F}^{k_F}dk\rho_1(k)\epsilon_0(k) =-\frac{\pi v_F}{6L^2}(1-\lambda).
\end{equation}
Substituting this into (\ref{GSE1}), we have 
\begin{equation}
\frac{E_{0,L}}{L}-\frac{E_0}{L} =-\frac{\pi v_F}{6L^2}.
\end{equation}
This confirms the result of (\ref{FREE}).

\section{Proof of $\varepsilon(T) =\varepsilon(0)+O(T^2/\nu)$}

In this appendix, we derive the low temperature expansions of $%
\varepsilon_c(k,T)$ and $\varepsilon_s(\lambda,T)$, by solving the integral
equations (\ref{DENZ}) by iteration.

At the zeroth-order we take 
\begin{equation}
\varepsilon^{(0)}_c(k,T) =\varepsilon^{(0)}_c(k),\hspace{.2in}
\varepsilon^{(0)}_s(\lambda,T) =\varepsilon^{(0)}_c(\lambda)=\frac{h}{2}.
\end{equation}
Then at the first order, i.e. after one iteration, the dressed energies are 
\begin{equation}
\varepsilon_c^{(1)}(k,T)=\varepsilon_c^{(0)} +O(Te^{-h/2T}),
\end{equation}
\begin{eqnarray}
\varepsilon_s^{(1)}(\lambda,T) &=&\varepsilon_s^{(0)}(\lambda)+\frac{1}{2\pi}
\int_{-k_F^{(0)}}^{k_F^{(0)}}dk K_1(\lambda-\sin k)\varepsilon_c^{(0)}(k) 
\nonumber \\
&&-\frac{1}{2\pi} \int_{-\lambda_F^{(0)}}^{\lambda_F^{(0)}}
d\lambda^{\prime}K_2(\lambda-\lambda^{\prime})
\varepsilon_s(\lambda^{\prime})  \nonumber \\
&&-\frac{T}{\pi}\int_{k_F^{(0)}-\delta}^{\pi} dk K_1(\lambda-\sin k)
\ln(1+e^{-\varepsilon_c^{(0)}})  \nonumber \\
&&+O(Te^{-h/2T}),  \label{ES1}
\end{eqnarray}
where $k_F^{(0)}$ is determined by $\varepsilon_c^{(0)}(k)=0$. We denote $%
\varepsilon_c^{(1)}(k)\equiv \varepsilon_c^{(0)}(k)$ and the first three
terms on the right hand side of (\ref{ES1}) by $\varepsilon_s^{(1)}(\lambda)$%
. The fourth term on the right hand side of (\ref{ES1}) is proportional to $%
T^2/v_c$, if we consider only the contributions from near the Fermi points.
Thus, we have 
\begin{eqnarray}
\varepsilon_c^{(1)}(k,T)&=& \varepsilon_c^{(1)}(k)+O(Te^{-h/2T}),  \nonumber
\\
\varepsilon_s^{(1)}(\lambda,T)&=& \varepsilon_c^{(1)}(\lambda)+O(T^2/v_c).
\end{eqnarray}
To get $\varepsilon_c^{(2)}(k,T)$ and $\varepsilon_s^{(2)}(k,T)$, it is
enough to replace $\varepsilon_c^{(1)}(k,T)$ and $\varepsilon_s^{(1)}(k,T)$
by $\varepsilon_c^{(1)}(k,0)$ and $\varepsilon_s^{(1)}(k,0)$ on the
iteration equations. Then by using similar techniques, we have 
\begin{eqnarray}
\varepsilon_c^{(2)}(k,T)&=& \varepsilon_c^{(2)}(k)+O(T^2/v_s),  \nonumber \\
\varepsilon_s^{(2)}(\lambda,T) &=&\varepsilon_c^{(2)}(\lambda)+O(T^2/\nu).
\end{eqnarray}
where $\varepsilon_c^{(2)}(k)$ and $\varepsilon_c^{(2)}(\lambda)$ are given
by the iteration of integral equations (\ref{DEZ}) and $\nu$ could be $v_s$
or $v_c$. Repeating the iteration, we finally have 
\begin{eqnarray}
\lim_{n\to\infty}\varepsilon_c^{(n)}(k,T)
&=&\lim_{n\to\infty}\varepsilon_c^{(n)}(k) +O(T^2/\nu),  \nonumber \\
\lim_{n\to\infty}\varepsilon_s^{(n)} (\lambda,k)&=& \lim_{n\to\infty}
\varepsilon_s^{(n)}(\lambda)+O(T^2/\nu).  \label{LIM}
\end{eqnarray}
Because of the convergence of the series for $\varepsilon_{c(s)}^{(n)}(k,T)$%
, (\ref{LIM}) implies that 
\begin{eqnarray}
\varepsilon_c(k,T)&=&\varepsilon_c(k) +O(T^2/\nu),  \nonumber \\
\varepsilon_s(\lambda,k)&=& \varepsilon_s(\lambda)+O(T^2/\nu),
\end{eqnarray}
which is what we intend to prove.

\section{An alternative proof of (\ref{BM})}

We give an alternative proof of (\ref{BM}). For the simplicity, we consider
the $h=0$ case in which, as Lieb and Wu pointed out, there are no
particle-like spinon excitations. The total energy is given by 
\begin{eqnarray}
\frac{E(T)}{L}&=&\int_{-\pi}^{\pi}dk \rho_c(k,T)\varepsilon^{(0)}_c(k,T) 
\nonumber \\
&+&\int_{-\infty}^{\infty}d\lambda \rho_s(\lambda,T)\varepsilon^{(0)}
_s(\lambda,T).  \label{TOT}
\end{eqnarray}
Substituting (\ref{DENZ}) into (\ref{TOT}), one has 
\begin{eqnarray}
\frac{E(T)}{L}&=& \int_{-\pi}^{\pi}dk \rho_c(k,T)\varepsilon_c(k,T) 
\nonumber \\
&+& \int_{-\pi}^{\pi}dkT\ln[1+e^{-\varepsilon_c(k,T)/T}] \frac{\cos k}{2\pi}
\nonumber \\
&\times&\int_{-\infty}^{\infty}d\lambda \rho_s(\lambda,T)K_1(\lambda-\sin k)
\nonumber \\
&+&\int_{-\infty}^{\infty}d\lambda \rho_s(\lambda,T)\varepsilon_s(\lambda,T)
\nonumber \\
&+&\int_{-\infty}^{\infty}d\lambda T\ln[1+e^{-\varepsilon_s(\lambda,T)/T}] 
\nonumber \\
&\times&\biggl [\frac{1}{2\pi}\int_{-\pi}^{\pi}dk
\rho_c(k,T)K_1(\lambda-\sin k)  \nonumber \\
&-&\frac{1}{2\pi}\int_{-\infty}^{\infty}d\lambda^{\prime}\rho_s(\lambda^{%
\prime},T)K_2(\lambda-\lambda^{\prime})\biggr].
\end{eqnarray}
Then divide the integration range of $k$ into the intervals $[-\pi, -k_F]$, $%
[-k_F,-k_F+\delta]$, $[-k_F+\delta, k_F-\delta]$, $[k_F-\delta,k_F]$ and $%
[k_F,\pi]$ and the interval of $\lambda$ into $[-\lambda_F,-\lambda_F+%
\delta] $, $[-\lambda_F+\delta,\lambda_F-\delta]$ and $[\lambda_F-\delta,%
\lambda_F]$. Use the fact that $e^{-|\varepsilon(k,T)|/T}$ decay rapidly as $%
T\to 0$ and (\ref{rho0}). After some algebras, one gets 
\begin{eqnarray}
\frac{E(T)}{L}&=&\frac{1}{2\pi} \int_{-k_F}^{k_F}dk\varepsilon_c(k,T) 
\nonumber \\
&&+2\int_{k_F-\delta}^{k_F+\delta}dk\rho_c(k_F) \frac{|\varepsilon_c(k)|} {%
1+e^{|\varepsilon_c(k)|/T}}  \nonumber \\
&&+2\int_{k_F-\delta}^{k_F+\delta}dk T\ln(1+e^{-|\varepsilon_c(k)|/T})
(\rho_c(k_F)-\frac{1}{2\pi})  \nonumber \\
&&+2\int_{\lambda_F-\delta}^{\lambda_F} d\lambda\rho_s(\lambda_F) \frac{%
|\varepsilon_s(\lambda)|} {1+e^{|\varepsilon_s(\lambda)|/T}}  \nonumber \\
&&+2\int_{\lambda_F-\delta}^{\lambda_F}d\lambda T
\ln(1+e^{-|\varepsilon_s(\lambda)|/T}) \rho_s(\lambda_F).
\end{eqnarray}
Here $\delta$ is a cut-off. Using the integral formulas 
\begin{equation}
\int_0^\infty dx\frac{x}{1+e^x} =\int_0^\infty dx\ln(1+e^{-x})=\frac{\pi^2}{%
12},
\end{equation}
we have, in the limit $T\to 0$, 
\begin{eqnarray}
\frac{E(T)}{L}&=&\frac{1}{2\pi} \int_{-k_F}^{k_F}dk\varepsilon_c(k)+\frac{1%
} {2\pi} \int_{-k_F}^{k_F}dk \tilde{\varepsilon}_c(k,T)  \nonumber \\
&&+\frac{\pi T^2}{6v_c}+\frac{\pi T^2}{6v_s} +\frac{\pi T^2}{6v_c}\biggl( 1-%
\frac{1}{2\pi\rho_c(k_F)}\biggr),  \label{B3}
\end{eqnarray}
On the other hand, at $h=0$, in terms of (\ref{Detzm}), we have an integral
equation for $\tilde{\varepsilon}_c(k,T)$, 
\begin{eqnarray}
&&\tilde{\varepsilon}_c(k,T) =-\frac{\pi T^2}{6v_c}\tilde{f}(k), \\
&&\tilde{f}(k)=\cos k_F \bar{K}(\sin k-\sin k_F)  \nonumber \\
&&+\frac{1}{2\pi}\int_{-k_F}^{k_F}dk^{\prime}\cos k^{\prime}\bar{K}(\sin
k-\sin k^{\prime})\tilde{f}(k^{\prime}).  \nonumber
\end{eqnarray}
Then comparing (\ref{rhoZM}) and $\displaystyle\int_{-k_F}^{k_F} \frac{dk}{%
2\pi}\tilde{\varepsilon}_c(k,T)$, one has 
\begin{equation}
\int_{-k_F}^{k_F}\frac{dk}{2\pi} \tilde{\varepsilon}_c(k,T)= -\frac{\pi T^2}{%
6v_c}\biggl( 1-\frac{1}{2\pi\rho_c(k_F)}\biggr).
\end{equation}
So 
\begin{equation}
\frac{E(T)}{L}=\frac{1}{2\pi} \int_{-k_F}^{k_F}dk\varepsilon_c(k) +\frac{\pi
T^2}{6v_c}+\frac{\pi T^2}{6v_s}
\end{equation}

Finally, using the thermodynamic relation 
\begin{equation}
E(T)=\frac{\partial(\beta\Omega(T))}{\partial \beta},
\end{equation}
we obtain the thermodynamic potential as given by (\ref{BM}).


\begin{references}
\bibitem{comm1}  This type of configurations can be realized by choosing $%
I_{N/2}=I_{N/2+1}=0$ in the ground state for $N_0=$even. For $N_0$=odd, the
configuration for the ground state is the same as usual.
\bibitem{comm3}  For $h\not{=}0$, $\delta E_0$ does not have a spin-charge
separated form. In this sense, the Virasoro algebras form a semi-direct
product. %%%%%%%%%%%%%%%%%%%%%
\bibitem{comm2}  Alternatively, we may extend the Fermi sea to that for two
separate right- and left-moving fermions with linear dispersion for $k$
covering the whole real axis in each sector. Then our commutation relations
are exact. (See D. C. Mattis and E. H. Lieb in \cite{TomMatt}.)
\end{references}
\end{document}